\documentclass[pre,superscriptaddress,twocolumn,showpacs]{revtex4} 
\usepackage{epsfig,amsmath,amssymb,graphics,color,calc} 
 
\newcommand{\be}{\begin{equation}} 
\newcommand{\ee}{\end{equation}} 
\newcommand{\ba}{\begin{eqnarray}} 
\newcommand{\ea}{\end{eqnarray}}

\begin{document} 

\title{Spontaneous and induced dynamic fluctuations in glass-formers I: \\ 
General results and dependence on ensemble and dynamics} 
 
\author{L.~Berthier} 
\affiliation{Laboratoire des Collo{\"\i}des, Verres
et Nanomat{\'e}riaux, UMR 5587, Universit{\'e} Montpellier II and CNRS,
34095 Montpellier, France}

\author{G.~Biroli} 
\affiliation{Service de Physique Th{{\'e}o}rique 
Orme des Merisiers -- CEA Saclay, 91191 Gif sur Yvette Cedex, France} 
 
\author{J.-P.~Bouchaud} 
\affiliation{Service de Physique de l'{\'E}tat Condens{\'e} 
Orme des Merisiers -- CEA Saclay, 91191 Gif sur Yvette Cedex, France} 
\affiliation{Science \& Finance, Capital Fund Management 
6-8 Bd Haussmann, 75009 Paris, France} 
 
\author{W. Kob} 
\affiliation{Laboratoire des Collo{\"\i}des, Verres
et Nanomat{\'e}riaux, UMR 5587, Universit{\'e} Montpellier II and CNRS,
34095 Montpellier, France}
 
\author{K. Miyazaki} 
\affiliation{Department of Chemistry, Columbia University, 3000 Broadway, 
New York, NY 10027, USA} 
 
\author{D. R. Reichman} 
\affiliation{Department of Chemistry, Columbia University, 3000 Broadway, 
New York, NY 10027, USA} 
 
\date{\today} 
 
\begin{abstract} 
We study theoretically and numerically a family of multi-point
dynamic susceptibilities that quantify the strength and characteristic
 lengthscales of dynamic heterogeneities in glass-forming materials. We
 use general theoretical arguments (fluctuation-dissipation relations
 and symmetries of relevant dynamical field theories) to relate the
 sensitivity of averaged two-time correlators to temperature
and density to spontaneous fluctuations of the local dynamics.  
Our theoretical results are then compared to molecular dynamics
 simulations of the Newtonian, Brownian and Monte-Carlo
 dynamics of two representative glass-forming liquids, a fragile binary
 Lennard-Jones mixture and a model for the strong glass-former silica.
We justify in detail the claim made in [Science, {\bf 310}, 1797 (2005)], 
that the temperature dependence of correlation functions allows one to extract 
useful information on 
dynamic lengthscales in glassy systems. We also discuss some subtle issues 
associated to the choice of
microscopic dynamics and of statistical ensemble through conserved
quantities, which are found to play an important role in determining 
dynamic correlations.
\end{abstract} 
 
 
% 64.70.Pf Glass transitions   
% 05.20.Jj  Statistical mechanics of classical fluids (see also
% 47.10.-g General theory in fluid dynamics)  
 
\maketitle 
 
\section{Introduction} 
 
Diverse materials, ranging from molten mixtures of metallic atoms, 
molecular and polymeric liquids, and colloidal suspensions may form 
glasses if sufficient undercooling or densification  
occurs~\cite{Donth,DS,walter}.  A glass 
may be characterized mechanically as a solid, but microscopically lacks 
the long-range order of a crystal.  Close to vitrification, the viscosity 
of glass-forming systems increases dramatically and sensitively as the 
thermodynamic control variables are changed.  Furthermore, some degree of 
universality is observed in the thermal and temporal behavior of systems 
close to the glass transition, even though the material properties of such 
systems may be vastly different~\cite{Donth,DS}.  
Despite decades of intense theoretical 
and experimental work, the underlying causes of this interesting behavior 
are not well understood. 
 
The observed quasi-universal behavior of glassy systems might be related to
the existence  
of a growing lengthscale as the glass transition is approached. 
The search for such a correlation lengthscale has led to 
intense activity in recent years. 
Static structural indicators have repeatedly failed to show any evidence
of collective behavior. 
Indeed, the static structure of a supercooled liquid hardly differs from
that of  
the same liquid at relatively high temperatures. Clearly all simple  
structural correlations remain short-ranged as the glass transition is  
approached~\cite{nostructure}.  It has become manifest in the 
last decade that interesting 
behavior is revealed by spatially correlated {\em dynamics}. 
As a whole, such effects are referred to as  
`dynamical heterogeneity'~\cite{ediger,sillescu,richert,sharon,hans}. 
 
The investigation, via 
theory~\cite{rfot,Gilles,jackle,FA,harrowell,gc,steve,BB},   
simulation~\cite{doliwa,harrowell2,yamamoto,japonais},  
and experiment~\cite{exp0,exp1,exp2,mark,encoremark,encoredonth},  
of various aspects of dynamic heterogeneity has  
greatly advanced our understanding of 
the behavior of systems close to the glass transition. In particular, 
multi-point susceptibilities have been devised to quantify the behavior 
and magnitude of the putative growing dynamical  
lengthscale~\cite{sharon,doliwa,yamamoto,FP,parisi,silvio2,glotzer,lacevic,gc,steve,berthier,TWBBB},
and experimental studies  
have, for several materials, directly determined the number of molecular
units that move cooperatively near the glass   
transition~\cite{exp0,exp1,mark,encoremark,mayer,dauchot,science}.   
Despite
recent breakthroughs, much more work needs to be done to fully 
characterize such a 
behavior both experimentally and theoretically. 
 
In the present work, contained here and in a companion paper~\cite{II}, 
we make a step towards this goal by investigating in detail different
susceptibilities  
that may be categorized according to the induced or spontaneous nature
of the measured fluctuations~\cite{science}. Spontaneous dynamic
fluctuations can be characterized by 
four-point functions, as proposed and studied 
earlier~\cite{sharon,doliwa,yamamoto,FP,parisi,silvio2,glotzer,lacevic,gc,steve,berthier,TWBBB}, or 
three-point functions, as in \cite{science,BBMR}.
Instead, fluctuations can be {\it induced} by monitoring
the change of dynamical correlators that follows a change of an external
control parameter, e.g. temperature \cite{science,BBMR}. 
As we shall show, it is possible
to relate induced and spontaneous dynamical 
fluctuations via fluctuation-dissipation relations as proposed in
\cite{science}. This 
provides a very valuable experimental tool to measure dynamic fluctuations 
since, as usual, induced fluctuations are much easier to measure than 
spontaneous ones. 
   
Using molecular dynamics simulations  of different archetypal
glass-forming liquids 
(e.g. ``strong'' materials that exhibit  an Arrhenius temperature
dependence of the viscosity, and ``fragile'' ones,  
whose viscosity displays a super-Arrhenius temperature dependence)
we shall show that 
in the slow dynamical regime a considerable fraction of spontaneous
fluctuations can be attributed to energy fluctuations: since the dynamics
is spatially correlated a local energy fluctuations induces a 
change in the dynamics over a much larger range.

Our analysis will however reveal that {\it global} 
four-point correlations describing the fluctuations of 
intensive dynamical correlators may depend both on the statistical ensemble 
and on the underlying microscopic dynamics. {\it Local} 
four-point correlations measuring the correlation between the relaxation
dynamics at a finite distance apart of course do not depend on the statistical
ensemble but they do depend on the underlying microscopic dynamics.
This is striking because it is known that correlators measuring the average dynamics do not depend in the relevant glassy regime on
the microscopic dynamics (Newtonian or 
stochastic~\cite{gleim,tobepisa}).
We address this problem both
theoretically and numerically, and conclude that, although the underlying physical mechanisms are the same, 
dynamical correlations depend quantitatively on the conserved physical quantities (and global
four-point correlations even on the statistical ensemble). 
For example, the absolute magnitude of global spontaneous dynamical
fluctuations in a Lennard-Jones system in the $NVT$ ensemble obtained from
Brownian dynamics (BD) or Monte-Carlo dynamics (MC) are 
very similar, but are considerably smaller than that obtained with
Newtonian dynamics (ND) in the same $NVT$ ensemble, 
whereas ND simulations performed in the $NVE$ ensemble yield results that 
are close to the BD and MC results in the $NVT$ ensemble. However, 
we stress that all our results point toward the conclusion that
the behaviour of all these quantities as the glass
transition is approached is governed by the growth of 
a {\it unique} dynamic correlation length, at least in the numerically accessible regime.
 
The aim of the present paper is to provide the reader with the physical
picture  underlying the dynamical susceptibilities introduced in \cite{science}, 
along with
more technical elements based on general field-theoretical considerations and  
detailed numerical investigations of different realistic 
glass-forming liquids. 
In a companion paper~\cite{II}, we present some quantitative 
predictions for these 
susceptibilities, obtained within different theoretical models:  
mean-field spin glass models~\cite{leticia}, mode-coupling
theory~\cite{Gotze},  and
kinetically constrained models~\cite{reviewkcm} which we  
again confront with the results from molecular dynamics simulations. 
The present paper is arranged in four   
sections. In Section~\ref{sectionjp} we  
present the physical motivations, definitions and physical 
content of several multi-point dynamic susceptibilities. We derive in
particular general results for the ensemble dependence of dynamic
fluctuations, fluctuation-dissipation relations, and bounds 
between induced and spontaneous dynamic fluctuations.  
In Section~\ref{sectiongiulioft}   
we present  a field-theoretic derivation of the behavior of dynamic  
fluctuations for various types of microscopic dynamics.  
This is particularly useful in identifying the precise physical mechanism leading 
to a growth of dynamic correlations, and the dependence of
multi-point susceptibilities on the microscopic dynamics. 
In Section~\ref{sectiongiulioend} we summarize our various theoretical predictions
and extract some important consequences, relevant to experiments, that need to be tested numerically.
In Section~\ref{MD} we present the results of detailed molecular dynamics  
simulations of two model glass-forming liquids,  
a fragile binary Lennard-Jones mixture and the strong 
BKS model for silica. We compare spontaneous and induced fluctuations and 
show that, as predicted theoretically, dynamic correlations 
strongly depend on the 
choice of microscopic dynamics and statistical ensemble. 
Our results suggest, however, that
a  unique dynamical lengthscale governs the growth of dynamical susceptibilities in all cases.
In Section~\ref{conclusion} we give the conclusions of our study. Although very natural in
spin-systems, four-point correlators in liquids mix dynamical heterogeneities with different physical 
effects (in particular, energy and density conservation) and might therefore not be the most effective object to work
with. On the other hand, we fully confirm the claim made in \cite{science}, that 
the temperature dependence of correlation functions allows one to extract rich and useful information on 
dynamic lengthscales in glassy systems \cite{exppaperinprep}.

\section{Multi-point dynamic correlators and new linear susceptibilities} 
 
\label{sectionjp} 
 
\subsection{Why four-point correlators? The spin glass case} 
 
No static correlation has yet been found to reveal any notable feature 
upon approaching the glass transition \cite{nostructure,Ernst1991}. 
Any lengthscale associated  
with the slowing down of the system must therefore be hidden in some dynamic  
correlation function. This issue is in fact deeply related to one  
of the most important question pertaining to the physics of disordered 
systems -- how can one define {\it long-range amorphous order} in such systems?

We know from the theory of spin glasses that the above oxymoron 
has in fact a precise answer: some hidden long-range order indeed 
develops at the spin glass transition~\cite{young}.  
In order to reveal this long-range order, 
conventional two-point functions are useless. Even if spins $s_{{\bf x}}$ 
and $s_{{\bf x}+{\bf y}}$ have 
non-zero static correlations $\langle s_{{\bf x}} s_{{\bf x}+{\bf y}} \rangle$ 
in the spin glass phase, the average over space for a given distance  
$y$ vanishes because the pairwise correlations randomly change sign 
whenever ${\bf x}$ changes.  
The insight of Edwards and Anderson is that one should first square 
$\langle s_{{\bf x}} s_{{\bf x}+{\bf y}} \rangle$  
before averaging over space~\cite{ea}.  
In this case, the resulting (four-spin) correlation 
function indeed develops long-range tails in the spin glass phase. 
This correlation in fact decays  
so slowly that its volume integral,  
related to the non-linear magnetic susceptibility of the material,  
diverges in the whole spin glass phase~\cite{nonlin}.  
 
The Edwards-Anderson idea can in fact be understood from a dynamical
point of view, which is important for understanding both the physics of
the spin glass just above the transition, and its generalization to
structural glasses.  
Consider, in the language of spins, the following four-point 
correlation function: 
\be 
S_4({\bf y},t) = [\langle s_{{\bf x}}(t=0) s_{{\bf x}+{\bf y}}(t=0) 
s_{{\bf x}}(t) s_{{\bf x}+{\bf y}}(t) \rangle]_x, 
\ee 
where the brackets $[...]_x$ indicate a spatial average. Suppose 
that spins $s_{{\bf x}}$ and $s_{{\bf x}+{\bf y}}$  
develop static correlations $\langle s_{{\bf x}} s_{{\bf x}+{\bf y}} 
\rangle$ within the glass phase. In this case, $S_4({\bf y},t \to \infty)$ 
will clearly  
converge to the spin glass correlation $[\langle s_{{\bf x}} 
s_{{\bf x}+{\bf y}} \rangle^2]_x$. More generally,  
$S_4({\bf y},t)$ for finite $t$ is able to detect {\it transient} 
tendencies to spin glass order, for example slightly above the spin glass  
transition temperature $T_c$. 
Close to the spin glass transition, both the persistence 
time and the dynamic length  
diverge in a critical way: 
\be  
S_4({\bf y},t) \approx y^{2-d-\eta} \hat 
S\left(\frac{y}{\xi},\frac{t}{\tau}\right),  
\ee 
where $\xi \sim (T-T_c)^{-\nu}$ 
and $\tau \sim (T-T_c)^{- z \nu}$. As mentioned above, the static 
non-linear susceptibility diverges as  
$\int d{\bf y} S_4({\bf y},t \to \infty) \sim \xi^{2-\eta}$. 
More generally, one can define a time-dependent dynamic  
susceptibility as: 
\be 
\chi_4(t) \equiv \int d{\bf y} \,\, S_4({\bf y},t), 
\ee 
which defines, provided $S_4({\bf 0},t) \sim 1$, a correlation volume, 
i.e. the typical number 
of spins correlated in dynamic events taking place  
over the time scale $t$. As we shall discuss below, $\chi_4(t)$ 
can also be interpreted as a quantitative measure of the  
dynamic fluctuations. Note however that the precise relation 
between $\chi_4$ and $\xi$ depends on the value of the 
exponent $\eta$, which is physically controlled by 
the detailed spatial structure of $S_4$: 
\be 
\chi_4(t=\tau) \propto \xi^{2-\eta}. 
\ee 
 
Therefore, spin glasses offer a precise example of a system which 
gets slower and slower upon approaching $T_c$ but  
without any detectable long-range order appearing in two-point 
correlation functions. Only more complicated  
four-point functions are sensitive to the genuine amorphous 
long-range order that sets in at $T_c$ and give 
non-trivial information even above $T_c$. In the case of 
spin glasses
it is well established that the transition is related to 
the emergence of a low temperature spin glass phase.
In the case of the glass transition of 
viscous liquids the situation is much less clear. 
There might be no true phase
transition toward a low temperature amorphous phase. 
It is still 
reasonable to expect that the dramatic increase of the 
relaxation time is due to 
a transient amorphous order that sets in and whose range 
grows approaching the glass transition.
Growing timescales should be somehow related to growing 
lengthscales~\cite{MS0}.
A good candidate to unveil the existence of this phenomenon is the
function $S_4({\bf y},t)$ introduced previously, 
since nothing in the above arguments was specific to systems with 
quenched disorder. 
The only difference 
is that although transient order is detected in $S_4({\bf y},t)$ 
or its volume integral $\chi_4(t)$ for times
of the order of the relaxation time, in the long time limit these 
two functions may not, and indeed do not in the
case of supercooled liquids, show long-range amorphous order. 
This roots back to the different nature of the 
glass and spin glass transitions (see the discussion in \cite{nonlin}). 

\subsection{Supercooled liquids and more multi-point correlations} 
 
In the case of liquids, we may consider a certain space 
dependent observable $o({\bf x},t)$, such as, for example, the 
local excess density $\delta \rho({\bf x},t)= \rho({\bf x},t) - \rho_0$, 
where $\rho_0$ is the average density of the liquid, 
or the local dipole moment, the excess energy, etc. 
We will assume in the following that the average of $o({\bf x},t)$ 
is equal to zero, and the variance of $o({\bf x},t)$ normalized to unity. 
The dynamic two-point correlation is defined as: 
\be \label{codef}
C_o({\bf r},t) =  [o({\bf x},t=0) o({\bf x}+ {\bf r},t)]_x, 
\ee 
where the normalization ensures that $C_o({\bf r}={\bf 0},t=0) = 1$. 
The Fourier transform of $C_o({\bf r},t)$ defines a generalized  
dynamic structure factor $S_o({\bf k},t)$~\cite{hansen}.  
All experimental and numerical results known to date suggest that as 
the glass transition 
is approached, no spatial anomaly of any kind appears 
in $C_o({\bf r},t)$ (or in $S_o({\bf k},t)$) although of course there  
could still be some signal which is perhaps too small to be measurable. 
The only remarkable feature is that the slowing down of the  
two-point correlation functions often obeys, 
to a good approximation, 
``time-temperature superposition'' in the $\alpha$-relaxation 
regime $t \sim \tau_\alpha$, i.e.: 
\be 
C_o({\bf r},t) \approx q_o({\bf r}) f\left(\frac{t}{\tau_\alpha(T)}\right), 
\ee 
where $q_o$ is often called the non-ergodicity (or Edwards-Anderson) 
parameter, and the scaling function $f(x)$ depends  
only weakly on temperature. This property will be used to simplify 
the following discussions, but it is not a crucial ingredient.
 
Whereas $C_o({\bf r},t)$ measures how, on average, the dynamics 
decorrelates the observable $o({\bf x},t)$, it is natural 
to ask whether this decorrelation process is homogeneous is space and
in time. Can the correlation last much  
longer than average? In other words, what is the distribution 
(over possible dynamical histories) of the correlation $C_o({\bf r},t)$? 
Clearly, since $C_o({\bf r},t)$ is defined as an average over 
some large volume $V$, the variance $\Sigma^2_C$  
of $C_o({\bf r},t)$ is expected to be of order $\xi^{2-\eta}/V$, 
where $\xi$ is the lengthscale over which $C_o({\bf r},t)$ is 
significantly correlated. 
More precisely
we define: 
\ba 
\Sigma^2_C & = & \int \frac{d{\bf x}}{V} \, \frac{d{{\bf x'}}}{V}   
o({\bf x},0) o({\bf x}+ {\bf r},t)  
o({{\bf x'}},0) o({{\bf x'}}+ {\bf r},t) \nonumber \\  
&  & - C_o({\bf r},t)^2, 
\ea 
which, using translational invariance, can be transformed 
into the space integral of a four-point correlation: 
\be 
\Sigma^2_C =  \int \frac{d{\bf y}}{V} S_4({\bf y},t), 
\ee 
where  
\ba\label{def4point} 
S_4({\bf y},t) & = & \bigg\{ [ o({\bf x},0)  
o({\bf x}+ {\bf r},t)  
o({\bf x} + {\bf y},0) o({\bf x} + {\bf y} + {\bf r},t) ]_x  \nonumber \\ 
& & -  [ o({\bf x},t=0) o({\bf x}+ {\bf r},t) ]_x^2 \bigg\}. 
\ea 
The variance of $C_o({\bf r},t)$ can thus be expressed as 
an integral over space of a four-point correlation 
function, which measures the spatial correlation of the 
temporal correlation. This integral over space is also  
the Fourier transform of $S_4({\bf y},t)$ with respect to ${\bf y}$  
at the wavevector ${\bf q}$ equal to zero. We want to insist  
at this stage that ${\bf r}$ and ${\bf y}$ in the above equations 
play very different roles: the former enters the very 
definition of the 
correlator we are interested in Eq.~(\ref{codef}), whereas the 
latter is associated with the scale over which the dynamics is  
potentially correlated. Correspondingly, great care will be 
devoted in the following to distinguish the wavevector ${\bf k}$, 
conjugate to ${\bf r}$, and ${\bf q}$ conjugate to ${\bf y}$. 
 
Specializing to the case ${\bf r}=0$ (local dynamics), one 
finally obtains~\footnote{A similar expression could be 
obtained when $C_o({\bf r},t)$ 
is computed using a time average instead of a space average; 
the resulting variance would now measure the temporal  
correlation of the temporal correlation.}: 
\be 
\Sigma^2_C \equiv \frac{\chi_4(t)}{N}. 
\ee 
The analogy with spin glasses developed above 
suggests that this quantity reveals 
the emergence of amorphous long-range order;  
it is in fact the natural diverging susceptibility in the context of $p$-spin  
descriptions of supercooled 
liquids, where a true dynamical phase transition occurs at a 
certain critical temperature~\cite{FP,KT,II,BB,BBMR}.  
Since in real systems no true 
phase transition is observed, one expects $\chi_4(t)$ to 
grow until $t \approx \tau_\alpha$ and decay back to zero  
thereafter. Until $\tau_\alpha$, there cannot be strong 
differences between a system with quenched disorder and a  
system where disorder is dynamically self-induced. 

However, contrary to spin glasses,
for which an underlying lattice structure exists, viscous 
liquids consist of molecules or atoms having 
continuum positions. As a consequence, one has to coarse-grain 
space in order to measure the fluctuations 
of the local relaxation dynamics. Local now means on a region of 
the order of the interparticle distance. 
Therefore, generically, $\chi_4(t)=V\Sigma^2_C$ correspond 
either to the fluctuations
of the Fourier transform of $C_o({\bf r},t)$ evaluated at a 
wave-vector, $k_0$, of the order of the 
first peak in the structure factor \cite{berthier}, or to a 
spatial average $\int d{\bf r} C_o({\bf r},t) w({\bf r})$ where 
$w ({\bf r})$ is an overlap function equal to one for lengths 
of the order of $2\pi/k_0$ and zero otherwise \cite{sharon}. 
The dependence of dynamical correlations on the coarse-graining 
length has been recently studied
in \cite{Ck} and is also discussed in the companion paper \cite{II}. 
 
Although readily accessible in numerical simulations, $\Sigma^2_C$ is 
in general very small and impossible to measure directly in 
experiments, except when the range of the dynamic correlation is 
macroscopic, 
as in granular materials~\cite{dauchot} or in
soft glassy materials where 
it can reach the micrometer and even millimeter range~\cite{mayer,luca}.
The central idea of this work is 
that induced dynamic fluctuations are more easily accessible than spontaneous 
ones, and can be related to one another by fluctuation-dissipation 
theorems. The physical motivation is that while four-point 
correlations offer a direct probe of the dynamic heterogeneities, 
other multi-point correlation functions give very useful information 
about the microscopic mechanisms leading to these heterogeneities. For 
example, one expects that the slow part of a local enthalpy 
(or energy, density) fluctuation per 
unit volume $\delta h$ at ${\bf x}$ 
and time $t=0$
triggers or eases the dynamics in its surroundings, leading to a 
systematic correlation between $\delta h({\bf x},t=0)$ and $o({{\bf 
x'}},t=0) o({{\bf x'}}+ {\bf r},\tau_\alpha)$. This defines a family of 
three-point correlation functions that relate thermodynamic or 
structural fluctuations to dynamics. Interestingly, some of these 
three-point correlations are both experimentally accessible and give 
bounds or approximations to the four-point dynamic correlations. The 
reason is as follows. In the same way that the space integral of the 
four-point correlation function is the variance of the two-point 
correlation, the space integral of the above three-point correlation 
is the covariance of the dynamic correlation with the energy 
fluctuations~\footnote{Note that for the enthalpy we use the 
notation $H(t=0)= \frac 1 N 
\int d{\bf x} h({{\bf x}},t=0)$.  
Therefore, $h$ is an enthalpy per unit volume.}:  
\ba & &\Sigma_{CH}= \frac{1}{VN} \int d{\bf x}\,d{{\bf 
x'}} o({\bf x'}+ {\bf r},t) o({\bf x'},0) \delta h({{\bf x}},0) 
\nonumber \\ & \equiv & \frac 1 N \int d{\bf y}\, [ o({\bf x}+ {\bf y} + {\bf 
r},t) o({\bf x} + {\bf y},0) \delta h({\bf x},0)]_x.   
\ea  
Hence, using the fact that the enthalpy fluctuations per particle are of order 
$\sqrt{c_P} k_B T$ (where $c_P$ is the specific heat in $k_B$ 
units), the quantity $N\Sigma_{CH}/\sqrt{c_P} k_BT$ defines the 
{\it number of particles} over which enthalpy and dynamics are correlated.
Of course, analogous 
identities can be derived for the covariance with density (and energy)
fluctuations. 
 
Now, on very general grounds, the covariance obeys the
Cauchy-Schwarz bound: $\Sigma_{CH}^2 \leq \Sigma_C^2 \Sigma_H^2$, 
where $\Sigma_H^2$ is the variance of the enthalpy fluctuations, 
equal to $c_P(k_B T)^2/N$ in the $NPT$ ensemble, $N=\rho_0 V$ being 
the total number of particles.  
Therefore, the dynamic susceptibility $\chi_4(t)$ is bounded from below by: 
\be \label{ineq} 
\chi_4(t) \equiv N \Sigma^2_C \geq 
\frac{N^2\Sigma_{CH}^2}{N\Sigma_{H}^2}=
\left(\frac{N\Sigma_{CH}}{\sqrt{c_P} (k_B T)}\right)^2, 
\ee 
where, as we show below, the right hand side can be
accessed experimentally. We then discuss in Sec.~\ref{lebovitz}  
how the above bound can be interpreted as an approximation, 
with corrections that can be physically estimated. Note that 
we chose here $\chi_4(t)$ to define a number of particles; 
one can of course convert it into a volume by multiplying   
$\chi_4$ by $v_0=1/\rho_0$, the average volume per particle. 
Note also that here and in the following 
we will work in the $NPT$ ensemble which is the relevant ensemble
for experiments 
on molecular liquids. We will discuss later the
generalization to different ensembles.
 
\subsection{A dynamic fluctuation-dissipation theorem and 
growing lengthscales} 
 
Consider a system in the grand-canonical $NPT$  
ensemble. The probability of a given configuration $\cal C$ is 
given by the Boltzmann weight $\exp( -\beta H[{\cal C}])/Z$, 
where $\beta=1/k_B T$ and $Z$ is the grand-partition function.
Suppose one  
studies an observable $O$ with the following properties: (i) $O$ only 
depends on the current microscopic 
configuration $\cal C$ of the system and (ii) $O$ can be written as 
a sum of local contributions:  
\be 
O = \frac{1}{V} \int d{\bf x}\; o({\bf x}). 
\ee 
In this case, a well-known static  
fluctuation-dissipation theorem holds~\cite{hansen}: 
\be 
\frac{\partial \langle O \rangle}{\partial \beta} = - 
\int d{\bf x} \, \langle o({\bf x}) \delta h({\bf 0}) \rangle  
\equiv - N\Sigma_{OH}, 
\ee 
where we decomposed the enthalpy in a sum of local  
contributions as well \cite{hansen}. 
 
Interestingly, in the case of {\it deterministic} 
Hamiltonian dynamics, the value of any local observable $o({\bf x},t)$ is 
in fact a highly complicated function of the initial 
configuration at time $t=0$. Therefore, the   
correlation function, now averaged over both space and 
initial conditions can be written as 
a thermodynamical average: 
\ba 
C_o({\bf r},t;T) = \frac{1}{Z(\beta)V}  
\int d{\bf x}\; o({\bf x}+{\bf r},t)o({\bf x},t=0) 
\nonumber \\ 
\times  
\exp\left[-\beta \int d{{\bf x'}}\; h({{\bf x'}},t=0)\right]. 
\ea 
Hence, the derivative of the correlation with respect to 
temperature (at fixed volume) directly leads, in the case of purely 
conservative Hamiltonian dynamics, to the covariance 
between initial energy fluctuations and the dynamical correlation. 
Defining $S_T({\bf x},t)=\langle o({\bf x}+{\bf r},t)o({\bf x},0)\delta  
h({\bf 0},0) \rangle$, one finds:
\be \label{fdt} 
 \frac{\partial C_o({\bf r},t;T)}{\partial T} =  \frac{1}{k_B T^2} 
\int d{\bf x}\; S_T({\bf x},t) \equiv \chi_T({\bf r},t). 
\ee
Hence, the sensitivity of the dynamics to temperature $\chi_T$ is directly  
related to a dynamic correlation. This last equality, although in a sense trivial, is one  
of the central result of this work. It has an immediate deep physical consequence, 
which is the {\it growth of a dynamical length upon cooling in glassy systems}, as we show now. 

Define $\tau_\alpha(T)$ such that $C_o({\bf
0},t=\tau_\alpha;T)=e^{-1}$ (say). Differentiating this definition
with respect to $T$ gives \be 0 = \frac{ d \tau_\alpha}{dT}
\frac{\partial C_o({\bf 0},t=\tau_\alpha;T)}{\partial t} +
\frac{\partial C_o({\bf 0},t=\tau_\alpha;T)}{\partial T}.  \ee Since
$C_o({\bf 0},t;T)$ decays from $1$ to zero over a time scale
$\tau_\alpha$, one finds that generically, using Eq. (\ref{fdt}): \be
\int d{\bf x}\; \frac{\langle o({\bf x},t=\tau_\alpha)o({\bf x},0)
\delta h({\bf 0},0) \rangle}{\rho_0 \sqrt{c_P} k_B T} \sim
\frac{1}{\rho_0 \sqrt{c_P}} \frac{d\ln \tau_\alpha}{d\ln T}.  \ee Now,
$\delta h$ is of order $\rho_0 \sqrt{c_P} k_B T$ and $\langle o^2
\rangle$ is normalized to unity, and the quantity $\chi_0 \equiv
S_T({\bf 0},\tau_\alpha)/\rho_0 \sqrt{c_P} k_B T$ cannot appreciably
exceed unity. The above integral can be written as $\chi_0 v_T$, which
defines a volume $v_T$ over which enthalpy fluctuations and dynamics
are appreciably correlated. Note that the interpretation of $v_T$ as a
true correlation volume requires that $\chi_0$ be of order one, and its 
increase is only significant if $\chi_0$ is essentially temperature independent.  
If this is not the case,
then the integral defined in (18) could grow due to a growing $\chi_0$
and not a growing length, which would obviate the notion that $v_T$
is a correlation volume.  For now we will assume these properties
hold, and return to this crucial point theoretically in more detail
in Sec.IV and with direct numerical evidence in Sec.V.

Assuming $\chi_0 \leq 1$, a divergence of the right hand side of 
the equality (\ref{fdt}) necessarily 
requires the growth of $v_T$. More precisely, as soon as $\tau_\alpha$ 
increases faster than any inverse power of temperature,  
the slowing down of a Hamiltonian system is necessarily 
accompanied by the growth of a dynamic correlation length. 
However, as already mentioned above, 
the precise relation between $\partial C/\partial \ln T$ 
and an actual {\it lengthscale}, $\xi$,  
depends on the value and structure of the spatial correlation 
function (for example the value of $\chi_0$ and 
the exponent $\eta$). In the simplest case of an exponentially 
decaying $S_T({\bf x},t)$, one finds:
\be
T \frac{\partial C_o({\bf r},t;T)}{\partial T} = 8 \pi \sqrt{c_V} 
\chi_0 \rho_0 \xi^3
\ee
 
It is instructive to study the case of a strong glass-former, for
which the slowing down is purely Arrhenius, i.e $\tau_\alpha=\tau_0
\exp[\Delta/(k_B T)]$, where $\Delta$ is some activation barrier. The
volume $v_T$ is then given by: \be v_T \sim \bigg| \frac{d\ln
\tau_\alpha}{d\ln T} \bigg| = \frac{\Delta}{k_B T}, \ee which
increases as the temperature is decreased, and diverges as $T \to 0$.
This is at first sight contrary to intuition since simple barrier
activation seems to be a purely local process. However, one should
remember that the dynamics strictly conserves energy, so that the
energy used to cross a barrier must be released from other parts of
the system. This release necessarily induces dynamic correlations
between the Arrheniusly relaxing objects. We conclude that even in a
strong glass the Arrhenius slowing down is necessarily accompanied by
the growth of a dynamic lengthscale. Note again that this conclusion
relies on subsidiary conditions that must be met.  Indeed, it is not
difficult to find examples of model Newtonian systems for which
$| d\ln \tau_\alpha / d\ln T|$ grows substantially
even though the physics is entirely local. In such cases, however, it
is expected that the spatial structure of $S_T({\bf x},t)$ will be
trivial, and the condition $\chi_0 \leq 1$ (independent of
temperature) will be violated.  In computer simulations, these
conditions may be checked, as we do in Sec.V.
  
\begin{figure*} 
\psfig{file=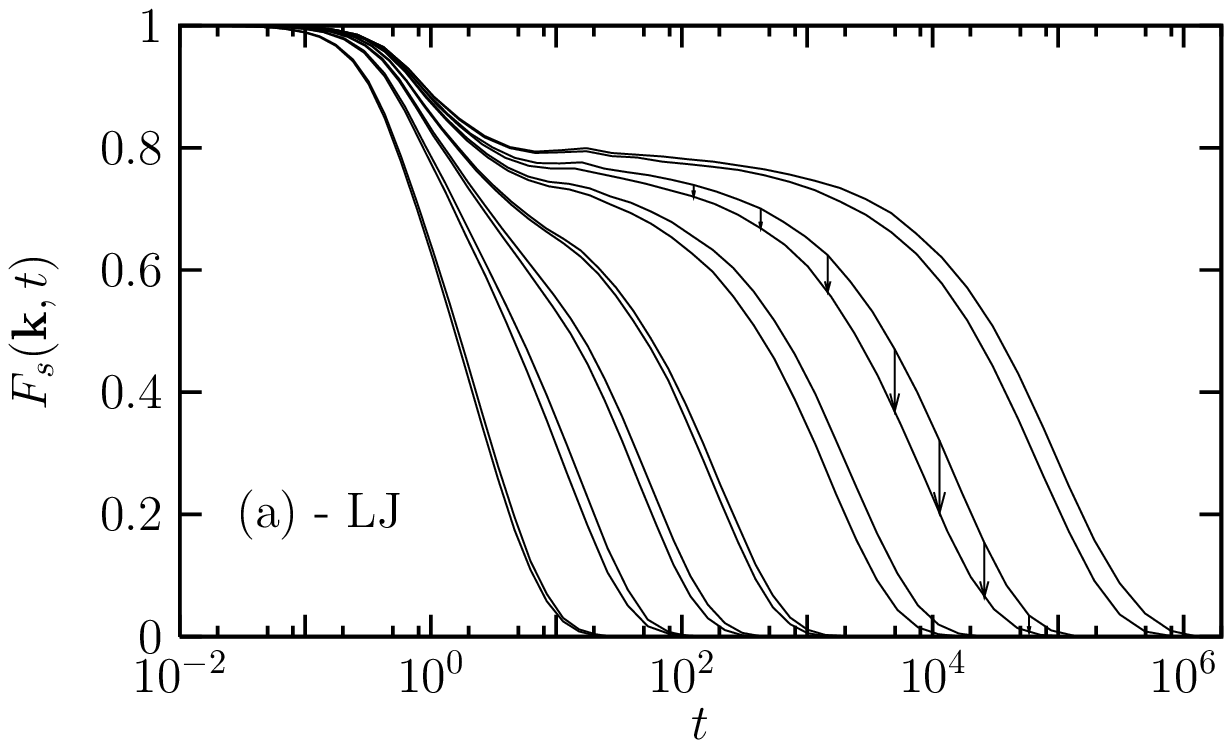,width=8.5cm} 
\psfig{file=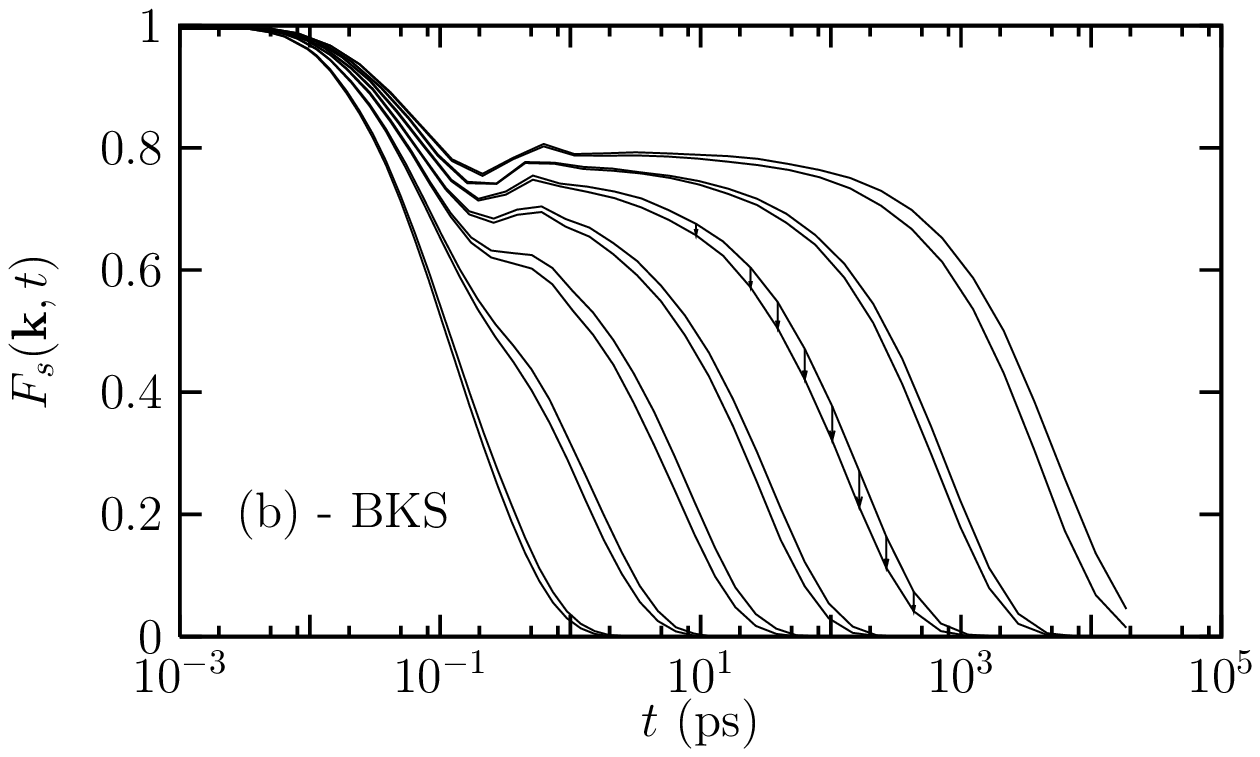,width=8.5cm} 
\psfig{file=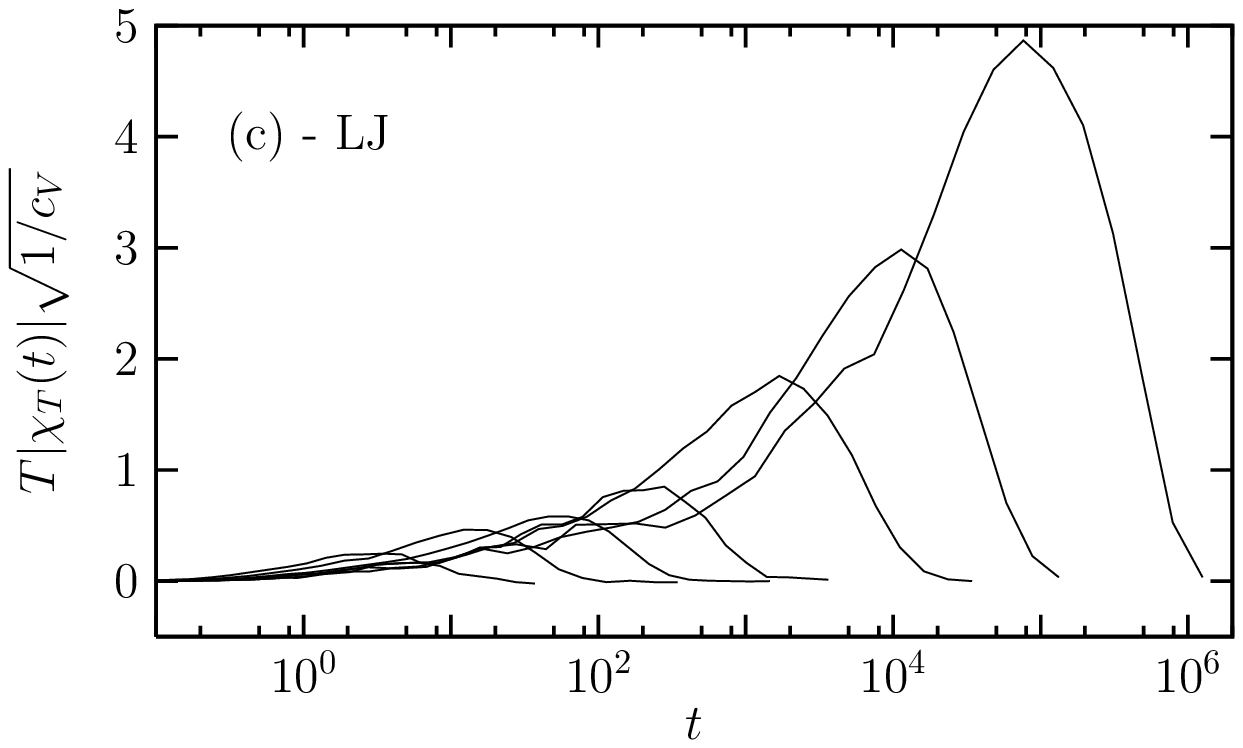,width=8.5cm} 
\psfig{file=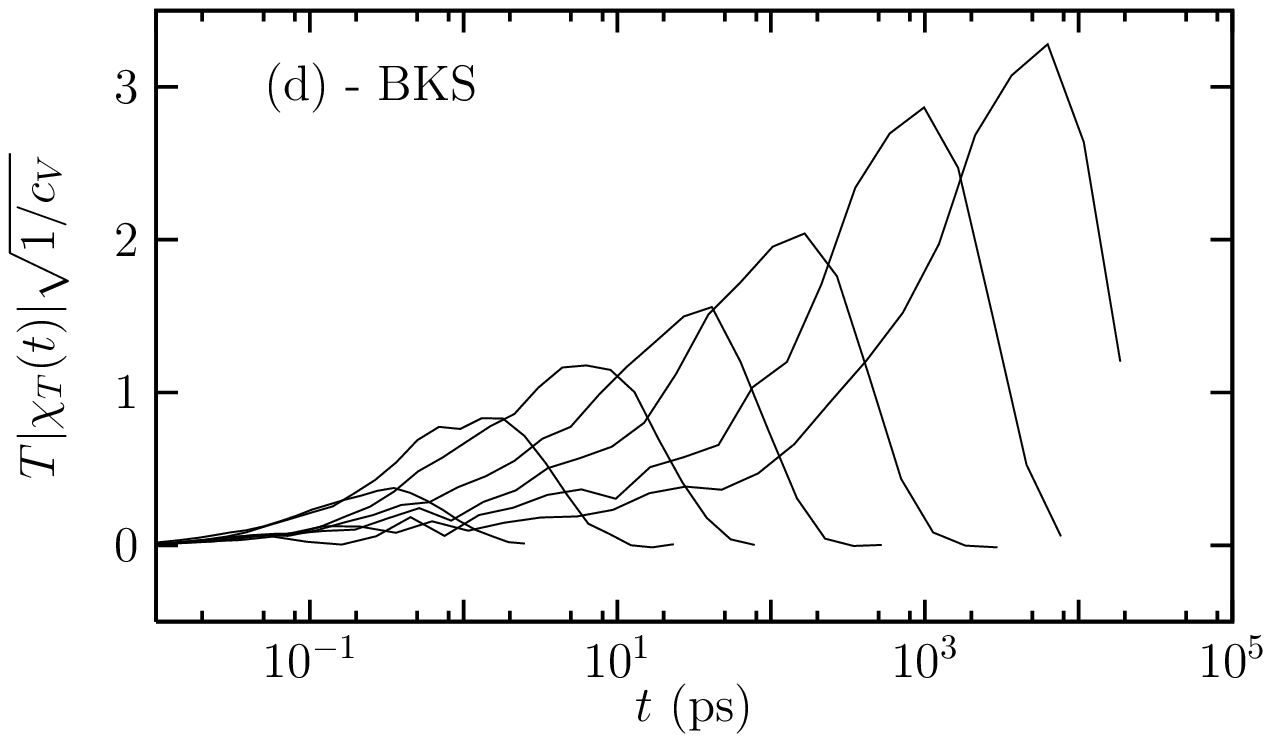,width=8.5cm} 
\caption{\label{figBKS1} (a) and (b) respectively show the  
self-intermediate scattering functions $F_s({\bf k},t)$  
as a function of time for various temperatures 
in a binary Lennard-Jones mixture and the BKS model for silica, 
obtained from the molecular dynamics numerical simulations  
discussed in Sec.~\ref{MD}. 
(a) $T=2.1$, 2.0, 1.05, 1.0, 0.75, 0.72, 0.61, 0.6, 0.51,  
0.5, 0.47, 0.46, 0.435 and 0.43 from left to right. 
(b) $T=6100$, 5900, 4700, 4600, 4000, 3920, 3580, 
3520, 3250, 3200, 3000, 2960, 2750, and 2715 K from left to right.  
The arrows illustrate how $\chi_T=\partial F_s / \partial T$  
is obtained by finite difference for each pair of temperatures.  
(c) and (d) show the resulting $\chi_T(t)$ for both models, normalized by the 
strength of energy fluctuations.   
We show the absolute value since $\chi_T$ is a negative quantity. 
In both liquids the dynamic susceptibility presents a peak for 
$t\approx \tau_\alpha$ whose height increases as temperature 
decreases, revealing increasingly 
heterogeneous and spatially correlated dynamics.} 
\end{figure*} 
 
When the relaxation time 
diverges in a Vogel-Fulcher manner, i.e. $\tau_\alpha=\tau_0 
\exp[DT_0/(T-T_0)]$, one finds that the corresponding  
dynamic correlation volume also diverges at $T_0$, as: 
\be 
v_T(\tau_\alpha) \sim  \frac{DTT_0}{(T-T_0)^2} \propto (\ln \tau_\alpha)^2, 
\ee 
where the last estimate holds sufficiently close to $T_0$. 
 
More generally, one can study the behavior of 
$\chi_T({\bf 0},t) \sim {\partial C_o({\bf 0},t;T)}/{\partial T}$ as a  
function of time. Since at all temperature $C_o({\bf 0},t=0;T)=1$ 
and $C_o({\bf 0},\infty;T)=0$, it is clear that  
$\chi_T({\bf 0},t)$ is zero at short and long times.  
We illustrate in Fig.~\ref{figBKS1}  
the shape of $\chi_T({\bf 0},t)$ for two glass-formers  
studied by molecular dynamics simulations described in Sec.~\ref{MD}. 
It has a peak for  
$t \approx \tau_\alpha$. It is a useful exercise to study the
example where the correlation 
function is a stretched exponential with exponent $\beta$ 
[not to be confused with $1/k_B T$], 
in which case: 
\be 
\frac{\partial C_o({\bf 0},t;T)}{\partial \ln T} = \frac{d\ln 
\tau_\alpha}{d\ln T} \, \, \beta \left(\frac{t}{\tau_\alpha}\right)^{\beta} 
\exp\left[-\left(\frac{t}{\tau_\alpha}\right)^{\beta}\right]. 
\ee 
This function behaves as a power-law, $t^\beta$,  
at small times and reaches a maximum for  
$t=\tau_\alpha$, before decaying to zero.  
The power-law at small times appears in the context of many  
different models, as discussed for the time behavior 
of $\chi_4(t)$~\cite{TWBBB}.  
Note also that for $t = \tau_\alpha$ and $T=T_g$, one has: 
\be 
\left.\frac{\partial C_o({\bf 0},\tau_\alpha;T)}{\partial 
\ln T}\right|_{T_g} =  \beta \, m \, \ln 10,  
\ee 
where $m=\left. Td\log_{10} \tau_\alpha /dT\right|_{T_g}$ is the 
steepness index, which characterizes the  
fragility of the glass.  
Note that in many cases, the resulting numerical value 
of $v_T \propto \chi_T$ turns out to be  
already large in the late $\beta$-regime, meaning that  
the concept of a cage is misleading because caging if fact 
involves the correlated motion of many particles~\cite{BBMR,dauchot}. 
 
Using the inequality in Eq.~(\ref{ineq}) with the results of 
the present section, we finally obtain a lower  
bound on the dynamical susceptibility $\chi_4(t)$ for Newtonian systems 
in the $NPT$ ensemble, which is 
experimentally accessible: 
\be \label{chi4bound} 
\chi_4({\bf r},t) \geq  \frac{T^2 \chi_T^2({\bf r},t)}{c_P} = \frac{1}{c_P} 
\left(\frac{\partial C_o({\bf r},t;T)}{\partial \ln T}\right)_P^2. 
\ee 
This bound implies that as soon as $\chi_T$ increases faster 
than $T^{-1}$ at low
temperatures, $\chi_4$ will eventually exceed unity; since $\chi_4$ 
is the space 
integral of a quantity bounded from above, this again means that the 
lengthscale
over which the four-point correlation $S_4(y,\tau_\alpha)$ extends 
{\it has to grow}
as the system gets slower and slower. More quantitative statements 
require information on
the amplitude and shape of $S_4(y,\tau_\alpha)$ which general 
field-theoretical and 
numerical results provide.

The above result in Eq.~(\ref{chi4bound}) is extremely general and 
applies to different situations discussed in the next 
section. It however does {\it not} apply when the dynamics is not 
Newtonian, as for instance for Brownian particles
or in Monte-Carlo numerical simulations~\cite{tobepisa,at,dh}.  
The reason is that in these cases, not only 
the initial probability but also the transition probability from the 
initial to the final configuration itself explicitly depends on temperature.  
In Brownian dynamics, for example, the noise in the Langevin equation 
depends on temperature~\cite{at}. 
Hence, ${\partial C_o({\bf r},t;T)}/{\partial T}$ 
receives extra contributions 
from the whole trajectory, that 
depend on the explicit choice of dynamics. We will argue below that when a 
dynamical critical point exists or is narrowly avoided, a system with 
Brownian dynamics should display dynamical correlations of the form 
$\chi_4 \sim \chi_T$ rather than the 
scaling $\chi_4 \sim \chi_T^2$ suggested by the above bound, 
Eq.~(\ref{chi4bound}). 
 
\subsection{Several generalizations} 
 
%\subsubsection{Enthalpy rather than energy} 
 
%The above results were obtained within the $NVT$  
%ensemble, where energy fluctuations are relevant. In many applications,  
%however, pressure rather than volume is held fixed. 
%In this fixed pressure ensemble, the probability of a given  
%configuration is given by: 
%\be\label{BolzH} 
%P({\cal C}) = \frac{1}{Z} \exp \left[-\beta \int d{\bf x}\;  
%h({\bf x}) \right], 
%\ee 
%where $h=e+\rho_0 Pv$ is the enthalpy density of the fluid, 
%$P$ is the pressure and $v$ the local volume per particle. Since the 
%variance of enthalpy fluctuations per particle is equal to  
%$N\Sigma^2_H= c_P k_B T^2$, all the above results carry through with 
%$c_V$ replaced by $c_P$, in particular: 
%\be \label{chi4boundNPT} 
%\chi_4({\bf r},t) \geq \frac{1}{c_P}  
%\left(\frac{\partial C_o({\bf r},t;T)}{\partial \ln T}\right)_P^2. 
%\ee 
 
\subsubsection{Density rather than temperature} 
 
In the above section, we have shown that the response of the 
correlator to a change of temperature is related to  
dynamic correlations. Other perturbing fields may also be 
relevant, such as density, pressure, concentration of  
species in the case of mixtures, etc. For example, for 
hard-sphere colloids, temperature plays very little 
role whereas small changes of density can lead to enormous 
changes in relaxation times~\cite{pusey}.  
Using the expression for  
the probability of initial configurations in the $NPT$  
ensemble, and the fact that the dynamics only depends on the 
initial condition, one now derives the following equality: 
\be  
\frac{\partial C_o({\bf r},t;P)}{\partial P}\bigg|_T = 
-\frac{\rho_0}{k_B T} \int d{\bf x}\; \langle 
o({\bf x}+{\bf r},t)o({\bf x},0)  
\delta v({\bf 0},0) \rangle, 
\ee 
which can again be used to define a dynamic 
correlation volume $\chi_\rho$. Introducing the isothermal compressibility  
$\kappa_T = (\partial \rho/\partial P)|_T/\rho_0$ 
and noting that the total variance of volume 
fluctuations per particle is given by  
$k_B T \kappa_T/ \rho_0$, we find: 
\ba \label{chirho} 
N\Sigma_{CV} & = &  
\rho_0 \int d{\bf x}\; \langle o({\bf x}+{\bf r},t)o({\bf x},t=0) 
\delta v({\bf 0},t=0) \rangle \nonumber \\ 
& = & -k_B T \kappa_T  
\frac{\partial C_o({\bf r},t;\rho)}{\partial \ln \rho}\bigg|_T, 
\ea 
from which we deduce  
a second bound on the dynamic correlation volume $\chi_4(t)$: 
\be \label{chi4boundrho} 
\chi_4({\bf r},t) \geq \rho_0 k_B T \kappa_T  \left(\frac{\partial 
C_o({\bf r},t;\rho)}{\partial \ln \rho}\right)_T^2. 
\ee 
Again, the right hand side of this expression is accessible to  
experiments~\cite{science}.  
Very importantly, and 
contrarily to the case of temperature,  
this inequality holds even for Brownian dynamics since the 
statistics of trajectories have no explicit dependence 
on pressure or density. Finally, a similar inequality holds for  
binary mixtures, relating $\chi_4(t)$  
to the dependence of the correlation function on the mixture composition.   
 
\subsubsection{Correlation and response in frequency space}

We have considered up to now the variance of the 
correlation function for a given time $t$, related to  
the four-point susceptibility $\chi_4(t)$, but 
this can be generalized to the covariance of the correlation
in frequency space. 

Defining $\hat C_o({\bf r},\omega)= \int_{0}^{\infty} dt \cos (\omega t) 
C_o({\bf r},t)$, the fluctuations of $\hat C_o({\bf r},\omega)$ 
define a four-point susceptibility in Fourier space  
$\chi_4({\bf r},\omega)$ given by: 
\be 
\chi_4({\bf r},\omega) = N\Sigma_{\hat C \hat C}
\ee 
Repeating the same argument developed for correlations in time space,
one finds: 
\be 
\chi_4({\bf r},\omega) = \frac{1}{c_P} \left(\frac{\partial 
C_o({\bf r},\omega;T)}{\partial \ln T}\right)^2. 
\ee 
We have up to now considered correlation functions, but the very 
same string of arguments also applies to linear  
response functions, which can, in the context of Newtonian dynamics, 
be written solely as functions of the initial 
condition. For example, the susceptibility of the observable $o$ to 
an external field $X$ is: 
\ba 
\chi_o({\bf r},t)  = & & \frac{1}{Z(\beta)V}    
\int  d{\bf x}\; \frac{\delta o({\bf x}+{\bf r},t)}{\delta X({\bf x},t=0)}  
\times \nonumber \\ 
 & & \exp\left[-\beta \int d{{\bf x'}}\; h({{\bf x'}},t=0)\right], 
\ea 
from which all the above results, transposed to response functions, 
can be derived. This is an important remark, since 
response functions, such as frequency dependent dielectric response 
or elastic moduli, are routinely measured in glassy  
materials. Their temperature or density dependence is therefore  
a direct probe of the dynamic correlation in these materials~\cite{science}. 
 
\subsubsection{Higher derivatives} 
 
One can of course study higher derivatives of the correlation 
functions with respect to temperature, which lead to 
higher order multi-point correlations between dynamics and 
energy or density fluctuations. For example, the second 
derivative gives a connected four-point correlation function: 
\be 
\frac{\partial^2 C_o({\bf 0},t;T)}{\partial \beta^2} = 
\int d{\bf x}\; d{\bf y}\; \langle o({\bf x},t)o({\bf x},0)\delta  
e({\bf y},0)\delta e({\bf 0},0) \rangle_c. 
\ee 
The right hand side now defines a squared correlation volume, 
where the left hand side, computed for $t = \tau_\alpha$, 
contains terms proportional to ${d^2\ln \tau_\alpha}/{d\ln T^2}$ 
and to $({d\ln \tau_\alpha}/{d\ln T})^2$. In most cases  
where $\ln \tau_\alpha$ diverges as an inverse power of temperature, or
in a Vogel-Fulcher-like manner, one finds that  
the latter term dominates over the former.  
This means that this squared correlation volume in  
fact behaves like $\chi_T^2$.  
The same argument also holds for higher derivatives.   
 
\subsection{Fluctuations and ensembles} 
\label{lebovitz} 
 
\subsubsection{Constrained vs. unconstrained fluctuations} 
\label{constrained} 
 
The above upper bounds in Eq.~(\ref{chi4bound})  
can in fact be given a much more precise meaning by  
realizing that fluctuations of thermodynamic quantities  
are Gaussian in the large volume limit~\cite{lebo},
except at a critical point.  
This allows one to show the following general  
result. Consider an observable $O$ that depends on  
$M$ Gaussian random variables $z_1,z_2,...,z_M$. We want 
to compare the ensemble where all the $z_i$'s are free to 
fluctuate with the ensemble where one constrains a subset of 
the $z_i$, say $z_{m},...,z_M$ to take fixed  
values, with no fluctuations. In the limit of small 
fluctuations, the variances of $O$ in the two ensembles are  
related through:  
\ba \label{general} 
&& \Sigma_O^2   =  \langle O^2 | z_{m},...,z_M \rangle_c + \nonumber \\ 
&& \sum_{\alpha,\beta=m}^M  
\frac{\partial \langle O | z_{m},...,z_M \rangle} 
{\partial z_\alpha} \frac{\partial \langle O | z_{m},...,z_M \rangle} 
{\partial z_\beta} \langle z_\alpha z_\beta \rangle_c, 
\ea 
where the average in the ensemble where $z_{m},...,z_M$ are 
fixed is denoted by $\langle \cdot |z_{m},...,z_M \rangle$.
The subscript $c$ means that we consider connected averages 
and we use Greek indices for the $(M-m+1)$ 
constrained variables. 
 
Because this result is important throughout this paper,
we sketch here its proof, using ideas   
and a notation which should make clear the analogy with a similar result
derived in Sec.~\ref{sectiongiulioft} using a field theoretical  
representation for the dynamics of supercooled liquids. Without loss of
generality, we can choose the mean of   
all $z_i$'s to be zero. The unconstrained joint distribution of the
$z_i$'s can be written as:  
\be 
P(\{z_i\}) = \frac{\sqrt{\det D}}{(2 \pi)^{M/2}} \exp \left( 
-\frac12 \sum_{ij} z_i D_{ij} z_j \right), 
\ee 
where $D$ is a certain $M \times M$ symmetric positive definite
matrix. The unconstrained covariance between $z_i$ and   
$z_j$ is well-known to be given by: 
\be 
\langle z_i z_j \rangle = (D^{-1})_{ij}. 
\ee 
Let us now write $D$ as blocks corresponding to the $(m-1)$ 
fluctuating variables and the $(M-m+1)$ fixed variables: 
\be  
D=\left[ 
\begin{array}{ccc} 
A & B\\ 
B^{\dagger} & C\end{array} \right], 
\ee 
where $A$ is $(m-1) \times (m-1)$, $B$ is $(m-1) \times (M-m+1)$ 
and $C$ is $(M-m+1) \times (M-m+1)$. When the  
variables  
$z_{m},...,z_M$ are fixed, the unconstrained variables acquire 
non-zero average values which are 
easily found to be given by: 
\be 
\overline{z_i} = \sum_{\alpha=m}^M (A^{-1}B)_{i \alpha} z_\alpha. 
\ee  
To establish the relation between constrained and unconstrained
covariances, we note the following block matrix inversion rule $D^{-1} =$:  
\be \label{matrixinversion} 
\left[ 
\begin{array}{cc} 
\{A-BC^{-1}B^{\dagger}\}^{-1} & -\{A-BC^{-1}B^{\dagger}\}^{-1}BC^{-1}\\ 
-C^{-1}B^{\dagger}\{A-BC^{-1}B^{\dagger}\}^{-1} & \{C-B^{\dagger}A^{-1}B\}^{-1}
\end{array} \right],  
\ee 
together with the matrix identity: 
\ba \label{magicrelation} 
\{A-BC^{-1}B^{\dagger}\}^{-1} & = &  
A^{-1}+ \\ 
& & (A^{-1}B)\{C-B^{\dagger}A^{-1}B\}^{-1}(A^{-1}B)^{\dagger}. 
\nonumber 
\ea 
The constrained covariance $\langle z_i z_j | z_{m},...,z_M \rangle_c$ 
is clearly given by $(A^{-1})_{ij}$. 
Using the above identities, we directly obtain: 
\be 
\langle z_i z_j \rangle \equiv \langle z_i z_j | z_{m},...,z_M \rangle_c +  
\sum_{\alpha, \beta}  
\frac{\partial \overline{z_i}}{\partial z_\alpha}  
\frac{\partial \overline{z_j}}{\partial z_\beta} 
\langle z_\alpha z_\beta \rangle . 
\ee 
Now, the final result Eq.~(\ref{general})  
above can be established simply by considering, to lowest order 
in the fluctuations, the observable $O$ as an $M+1$th
Gaussian variable correlated with all the $z_i$'s and apply the 
above equality to $i=j=M+1$.

\subsubsection{From $NPH$ to $NPT$} 
 
Let us apply the general result Eq. (\ref{general}) to the  
case of interest here, first to the case $M=1$, with $z_1=H$ and 
number of particles fixed. The two 
ensembles correspond to $NPH$ and $NPT$, respectively.  
The above formula can be used with the correlation $C_o$ as 
an observable provided the dynamics is conservative,  
as argued above. Therefore: 
\be 
\chi_4^{NPT}({\bf r},t) = \chi_4^{NPH}({\bf r},t) + \frac{1}{c_P} 
\left(\frac{\partial C_o({\bf r},t;T)}{\partial \ln T}\right)_P^2, 
\ee 
where we have replaced in the second term in the right hand side 
$\partial/\partial H$ by $(1/N c_P k_B) \partial/\partial T$; 
$\chi_4^{NPH}({\bf r},t)$ is the variance of the correlation  
function in the $NPH$ ensemble where enthalpy does not fluctuate, a 
manifestly non-negative quantity.  
Therefore, the above equation recovers the lower bound Eq. (\ref{chi4bound}), 
with a physically explicit expression for the missing piece.  
The relative contribution of the two terms determining  
$\chi_4^{NPT}$ will be discussed in concrete cases in  
Secs.~\ref{sectiongiulioft} and \ref{MD}.  
 
\subsubsection{Local vs. global fluctuations} 
 
The above discussion may appear puzzling for the 
following reason: we have seen that the susceptibility  
$\chi_4(t)$  
is the space integral of a four-point correlation function 
$S_4({\bf y},t)$ which, although developing some spatial  
correlations on approaching the glass transition, remains 
relatively short-range 
in the supercooled liquid phase and should {\it not} depend 
on far away boundary  
conditions that ultimately decide whether energy is  
conserved or not. 
Since $S_4({\bf y},t)$ does not depend, in the 
thermodynamic limit, on the ensemble, how can its integral over 
space, $\chi_4(t)$, be affected by the choice of ensemble? The  
answer is that while the finite volume corrections to $S_4({\bf y},t)$ 
for a given ${\bf y}$ tend to zero when  
$V \to \infty$,  the integral over space of these corrections  
remain finite in that limit~\cite{lebo}, and 
explain the difference between $\chi_4^{NPT}$ and $\chi_4^{NPH}$.  
We understand that the physical correlation  
volume is given by $\chi_4^{NPT}$; the long-range nature of the fixed 
energy constraint leads to an underestimate 
of $\chi_4$ in the $NPH$ ensemble, which is irrelevant to describe local  
correlations. This is particularly important in  
numerical simulations~\cite{lebo}: the study of $S_4({\bf q},t)$
(the Fourier  
transform of $S_4({\bf y},t)$) in the microcanonical 
ensemble will lead to a singular behavior associated to  
the fact that $\lim_{q \to 0} S_4({\bf q},t) \neq 
S_4({\bf q}=0,t)$, whereas the two coincide only in the 
ensemble where all conserved quantities are free to 
fluctuate ($NPH$ for monoatomic liquids). 
The former quantity is the physical quantity independent of  
the ensemble and will be denoted $\lim_{q \to 0} S_4({\bf q},t) 
= \chi_4^{*}$ in the 
following, whereas the 
latter depends on the macroscopic constraint. We summarize this important
discussion in Sec.~\ref{sectiongiulioend}.
 
\subsubsection{Various sources of fluctuations} 
 \label{gf}
Equation (\ref{general}) makes precise the intuition that dynamic 
fluctuations are partly induced by the fluctuations  
of quantities that physically affect the  
dynamic behavior~\cite{Donth,edigertheo}.  
Among these quantities, some are conserved thermodynamic quantities, such  
as the energy or density, and the dependence of the dynamics 
on those quantities are simply measured by the derivatives 
of the correlation function. The contribution of the local 
fluctuations of these quantities can therefore be estimated  
and lead to a lower bound to the total dynamic fluctuations. 
In a supercooled liquid one expects on general grounds that 
energy and density should play major roles  
in the dynamics. From the thermodynamic theory of  
fluctuations~\cite{landau}, we know 
that in fact temperature (seen formally as a function of 
energy and density) and density are independent random variables,  
with variance $\langle \delta T^2 \rangle= T^2/(N c_V)$ and  
$\langle \delta v^2 \rangle=k_B T \kappa_T/(N \rho_0)$. Therefore   
Eq.~(\ref{general})  
gives for the ``true'' dynamic susceptibility: 
\be \label{landau} 
\chi_4^{*} = \frac{1}{c_V} \left(\frac{\partial C_o}{\partial 
\ln T}\right)_V^2  
+  
\rho_0 k_B T \kappa_T   
\left(\frac{\partial C_o}{\partial \ln \rho}\right)_T^2 + \chi_4^{NVE}, 
\ee 
The question of whether other, ``hidden'' variables  
also contribute to the dynamic fluctuations is tantamount to  
comparing $\chi_4^{NVE}$ with $\chi_4^{*}$.  
This question is very difficult to resolve theoretically in general.  
The rest of this paper and the companion paper~\cite{II}  
are devoted to theoretical arguments and numerical simulations  
which attempt to clarify this issue. Our numerical results 
suggest that $\chi_4^{NVE} \ll \chi_4^{*}$, at least close to the glass 
transition, but 
that both $\chi_4^{NVE}$ and $\chi_4^{*}$ are in fact governed by the 
very same physical 
mechanism and define the same dynamical correlation length. 

Whether energy or density fluctuations is the dominant factor can be assessed 
by comparing the two explicit terms appearing in the right  
hand side of Eq. (\ref{landau}). Assuming time-temperature 
superposition, the ratio $r$ of the two terms for $t=\tau_\alpha$ 
reads: 
\be 
r = \rho_0 c_V k_B T \kappa_T 
\left(\frac{\frac{d\ln \tau_\alpha}{d\ln \rho}|_T}{\frac{d\ln 
\tau_\alpha}{d\ln T}|_\rho}\right)^2. 
\ee 
Following Ref.~\cite{alba},  
and noting that $\rho_0 c_V k_B T \kappa_T < 1$  
in usual liquids, we conclude that for 
most glass-formers, $r$ is significantly less than one, 
which means that density effects are weaker than temperature 
effects and consequently contribute little to dynamic fluctuations. 
The situation is of course completely the 
opposite in hard-sphere colloidal glasses, where ${d\ln \tau_\alpha} 
/{d\ln T}|_\rho \to 0$ and $r \gg 1$. 
 
\subsection{Summary} 
 
After motivating the use of multi-point correlation functions 
to detect non-trivial dynamic correlations in amorphous  
materials, we discussed the idea that induced fluctuations are 
more easily accessible experimentally  
than spontaneous ones, and can be related to one another by 
fluctuation-dissipation theorems. Elaborating on this idea,  
we have shown 
that the derivative of the correlation function with respect 
to temperature or density directly gives access to the volume integral of the 
correlation between local energy (or density fluctuations) 
and dynamics. This relation can be used to show on very 
general grounds that a sufficiently abrupt slowing down of the 
dynamics must be accompanied by the growth of a correlation volume. 
The detailed relation between these susceptibilities and  a 
correlation lengthscale
however depends on the amplitude and spatial structure of the 
multi-point correlation functions. 
 
We have then shown that the dynamic four-point susceptibility at $q=0$, 
which corresponds to the fluctuation of global intensive dynamical correlators,
depends in general on the chosen statistical  
ensemble. In the case where conserved variables are allowed to fluctuate, 
we showed that the dynamic  
four-point susceptibility is bounded from below by terms that capture 
the contribution 
of energy and density fluctuations to dynamic heterogeneities. Our 
central results, suggesting a way to estimate a dynamic 
correlation volume from experiments, are given in
Eqs. (\ref{chi4bound}, \ref{landau}).  
Whereas we expect that for most supercooled liquids, the contribution of 
temperature 
is the dominant effect, the quality of our bounds as  
quantitative estimators of $\chi_4$, and their physical relevance 
is, at this stage of the discussion, an open question  
which we carefully address below, in particular in 
Sec.~\ref{sectiongiulioend}, and in the companion  
paper~\cite{II}. The following section is devoted to a quantitative 
study of this question within a field-theory formalism. A surprising 
outcome of this analysis is that 
the dynamic four-point susceptibility at $q=0$
correlations not only depend on the chosen statistical 
ensemble, as shown above, 
but also on the choice of microscopic dynamics whether 
Newtonian or stochastic. Of course 
the dynamic four-point susceptibility at non-zero $q$
depends only on the choice of microscopic dynamics.
 
\section{Correlation of dynamical fluctuations: 
a field-theoretical perspective} 
 
\label{sectiongiulioft} 
 
In the following we develop in detail an approach to dynamical 
fluctuations in supercooled liquids based on general field-theory techniques, 
and discuss how a non-trivial lengthscale can be generated by interactions
and manifests itself in quantities like $\chi_4$ or $\chi_T$. 
We identify precisely 
the `susceptibility' (called $A^{-1}$ below) responsible for all 
interesting dynamic correlations. 
We discuss the origin of the ensemble dependence of dynamic 
fluctuations described above from a 
diagrammatic point of view. This is important since any 
self-consistent resummation or approximation scheme must be 
compatible with the bounds derived above.  
This formalism furthermore predicts that, contrary to the behavior of
correlators measuring the average dynamics, the details of dynamic
fluctuations depend on the dynamics in a remarkable way. However, since in all cases,
the object responsible for the increase of these dynamic correlations is the 
very same susceptibility $A^{-1}$, the physics revealed by the 
correlations is independent both of the ensemble and of the dynamics, and genuinely reflects the 
collective nature of glassy dynamics. 
  
In the companion paper \cite{II} we will point out 
how simplifications can occur if 
a true dynamical critical point exists, as 
within mode-coupling theory, a particular self-consistent resummation scheme.  
In the following we aim instead at keeping the discussion more general 
than the confines of mode-coupling theory or any other particular theoretical approach. This is important 
since mode-coupling theory is not expected to apply 
close to the glass transition temperature, whereas the present 
physical conclusions do.
 
\subsection{The dynamic field-theory} 
 
\subsubsection{A reminder of the usual static case} 
 
The dynamic field-theory strategy is analogous  
to the one used for ordinary static critical phenomena which we now recall, 
focusing on the ferromagnetic Ising  
transition as a pedagogical example~\cite{Zinn-Justin}.  
The starting point is the Legendre functional 
transform $\Gamma(m({\bf x}))$ of the free energy $\beta F(h({\bf x}))$,  
itself defined as a functional of  
the magnetic field $h({\bf x})$: 
\begin{equation} 
\Gamma(m({\bf x}))=\beta F(h({\bf x}))-\int d{\bf x'} h({\bf x'})m({\bf x'}), 
\end{equation} 
where $h({\bf x})$ on the right hand side is the field that 
leads to the magnetization profile $m({\bf x})$.  
The magnetization is determined via the equation: 
\begin{equation} 
m({\bf x})=\frac{\delta \beta F}{\delta h({\bf x})}.
\end{equation} 
Two important properties of the functional $\Gamma(m({\bf x}))$ 
that can be directly derived using the previous 
relation are: 
\ba 
\frac{\delta \Gamma}{\delta m({\bf x})}&=&-h({\bf x}), 
\nonumber \\  
\frac{\delta^2\Gamma}{\delta m({\bf x})\delta m({\bf x'})} & = & 
\frac{\delta h({\bf x})}{\delta m({\bf x'})}\equiv [\langle s({\bf x})  
s({\bf x'})\rangle]^{-1}.  
\ea 
The last exact identity indicates that the operator obtained by differentiating the functional $\Gamma$ twice is the  
inverse of the spin-spin correlation function (considered as an operator). Note that these are simple
generalizations of usual thermodynamic relations.   
 
In general one cannot compute $\Gamma$ exactly, but one can guess its form using symmetry  
arguments, and compute it approximately in a perturbative (diagrammatic) expansion in some parameter.  
Using the above identities, no further approximation is needed to obtain correlation functions.  
In its simplest version, $\Gamma$ corresponds to the Ginzburg-Landau 
free energy functional. The saddle point equation for  
the magnetization then leads to the mean-field  
description of the transition, whereas the second derivative term gives   
the mean-field result 
for the spin-spin correlation function, valid when the space dimensionality 
is sufficiently large. 
 
In the following, we will present a theory of dynamic fluctuations 
within a field-theoretic framework similar  
to the above static formalism. The main difference is that in the context of glassy dynamics, the relevant order 
parameter is no longer a one-point function like the magnetization  
but instead a two-point dynamic function which has to be introduced 
as an effective degree of freedom in the dynamic free-energy functional.  
 
\subsubsection{Dynamic free-energy functionals and fluctuations} 
 
Different dynamic field-theories have been used in the literature  
to analyze the dynamics of dense liquids. The common strategy is to 
write down exact or  
phenomenological stochastic equations for the evolution of the slow conserved 
degrees of freedom. For instance, for Brownian dynamics the only conserved 
quantity is the local density (energy and momentum are not 
conserved). The equation for the  
local density is in that case the so-called Dean-Kawasaki 
equation~\cite{Dean, Kawa}, which can be derived exactly for Langevin 
particles (see Refs.~\cite{ABL,das} for a discussion of different 
field-theories associated with such dynamics).  
In general, the field-theory associated to a given stochastic dynamics 
is obtained through the  
Martin-Siggia-Rose-deDominicis-Janssen method, 
where one first introduces response fields enforcing the correct time
evolution and then averages over the stochastic 
noise~\cite{Zinn-Justin,ABL}.
 
We will use a general notation that will allow us to treat all field  
theories proposed   
in the literature \cite{ABL,das} on the same   
footing. In all those field theories one has a set of slow
conserved fields, $\phi_i$ ($i=1,...,m$),   
and the corresponding response fields, $\hat \phi_i$ arising from the
Martin-Siggia-Rose   
procedure~\cite{Cardy}. It will also be useful to put $\phi_i, \hat
\phi_i$ into a single $2m$ dimensional vector $\Phi_a$,
$a=1,...,2m$. The average over the dynamic action of $\Phi_a$   
will be denoted $\Psi_a$: $\langle  \Phi_a \rangle= \Psi_a$.  
As in the static case, the starting point of the analysis is   
a Legendre functional (also called the generator of two-particle
irreducible  diagrams or Baym-Kadanoff functional)
\cite{Zinn-Justin,DeDominicisMartin,BlaizotRipka}.
It is equal to: 
\begin{widetext} 
\begin{align} 
& 
\Gamma(\Psi_a,G_{a,b})=-\ln \int {\cal D}\Phi_a \,  \exp\left(-S(\{\Phi_a\}) 
-\int dtd{\bf x} \sum_{a=1}^{2n} h_a({\bf x},t)[ 
\Phi_a({\bf x},t)-\Psi_a({\bf x},t)]\right.\\ 
& 
\left.-\frac 1 2 \int dtdt'd{\bf x}d{\bf x}' \sum_{a,b=1}^{2n}
 K_{a,b}({\bf x},t;{\bf x}',t') 
[\Phi_a({\bf x},t)\Phi_b({\bf x}',t')-\Psi_a({\bf x},t)
\Psi_b({\bf x}',t')-G_{a,b}({\bf x},t;{\bf x}',t')] \right), 
\end{align} 
\end{widetext} 
where $S$ is the action of the field theory,
$h_a$'s are such that $\langle  \Phi_a \rangle= \Psi_a$ and
$K_{a,b}$ imposes a certain value for the two-point functions:
$\langle\Phi_a \Phi_b\rangle  
-\Psi_a\Psi_b=G_{a,b}$. The properties of $\Gamma(\Psi_a,G_{a,b})$ are
the same as in the static  
case because formally it is the same mathematical object. The only
difference is that the dynamical   
functional depends on a larger number of variables.   
The difficulty is to devise an approximate expression for  
the functional $\Gamma$. Once this is done,  
one should differentiate the functional 
once to obtain self-consistent equations 
for the order parameters $\Psi_a,G_{a,b}$ and twice to obtain  
(after inversion) and expression for their fluctuations. More precisely,  
we introduce the following matrix of second derivatives: 
\[ 
\partial^2 \Gamma = \left[ 
\begin{array}{ccc} 
\frac{\delta \Gamma}{\delta G_{a,b}\delta G_{c,d}} & \frac{\delta 
  \Gamma}{\delta G_{a,b} \delta \Psi_e}\\ 
\frac{\delta 
  \Gamma}{\delta \Psi_f \delta G_{a,b}} & \frac{\delta 
  \Gamma}{\delta \Psi_e \delta \Psi_f }\end{array} \right] 
\equiv  
\left[ 
\begin{array}{ccc} 
A & B\\ 
B^{\dagger} & C\end{array} \right],   
\] 
where we have introduced three block matrices $A,B,C$,  
in full correspondence with those 
introduced above in  
Sec.~\ref{constrained}. The inversion of $\partial^2 \Gamma$ 
allows one to obtain the objects of interest in this paper. For example, 
inversion in the ``$GG$-sector'' defines the four-point space-time correlation functions: 
\begin{align} 
& (\partial^2 \Gamma)^{-1,G}_{a,b;c,d}  =  \langle \left( 
\Phi_a({\bf x},t)\Phi_b({\bf x}',t')-\Psi_a({\bf x},t)\Psi_b({\bf 
  x}',t')\right)  \nonumber \\ 
&  \,\,\, \times \left( 
\Phi_c({\bf y},s)\Phi_d({\bf y}',s')-\Psi_c({\bf y},s)\Psi_d({\bf 
  y}',s')\right)\rangle_c , 
\end{align} 
where $\langle \cdot \rangle_c$ means that we are focusing on  
the connected component. Similarly, inversion in the ``$G\Psi$-sector''
defines the three-point functions, such as the energy-correlation correlator defined in the 
previous section, whereas inversion in the ``$\Psi\Psi$-sector'' leads to
the exact propagators of the conserved quantity. For example, when $\Psi$ is the energy, one
obtains the exact energy propagator (dressed by interactions), which is expected to be 
diffusive in the hydrodynamic limit. 
 
At this stage, it is important to recall that the dynamical functional
$\Gamma$ has a direct diagrammatic expression   
as~\cite{Zinn-Justin,DeDominicisMartin}: 

\begin{equation}\label{2PI} 
\Gamma(\Psi,G)=-\frac{1}{2}\mbox{Tr} \log G 
+\frac{1}{2}\mbox{Tr}\, G_{0}^{-1} [G+\Psi\Psi]-
\Phi_{2PI}(\Psi,G), 
\end{equation} 
where $\Phi_{2PI} (\Psi,G)$ is the sum of all two  
particle irreducible Feynman diagrams (that cannot be decomposed in two  
disjoint pieces by cutting two lines) constructed with the vertices of 
the theory and using the full propagator $G$ as lines and $\Psi$ as  
sources~\cite{Zinn-Justin,DeDominicisMartin,BlaizotRipka}. 
Both the internal indices
and spatio-temporal arguments were skipped for simplicity.   
The first derivatives lead to the self-consistent equations for the
order parameter. Since in dynamical field-theories for liquids the slow
physical fields   
are in fact conserved quantities the equations $\delta
\Gamma/\delta\Psi_a=0$ do not fix the values of the physical fields  
that have to be fixed by the initial 
conditions. On the other hand, they set to zero the average of the
response fields and   
enforce translational invariance \footnote{Note that since the value of
the physical fields is fixed by the initial conditions and not changed
by loop corrections, it is often more useful and practical to develop the
theory in terms of $\delta \Psi$ so that the average of the fluctuating
fields is zero by construction.}.   
 
On the other hand, the derivatives $\delta \Gamma/\delta G=0$ lead to   
formally exact self-consistent 
equations for the two-point  
correlation functions. These equations can be written as a   
Schwinger-Dyson matrix equation: 
\[ 
G^{-1}=G_0^{-1}-\Sigma(G),\qquad \Sigma(G)=\frac{\delta
\Phi_{2PI}}{\delta G},  
\] 
where $\Sigma$ is the self-energy. A given approximation consists in
retaining a given set  
of diagrams in $\Phi_{2PI}$, or alternatively in $\Sigma(G)$.   
For example, mode-coupling  
theories generically consist in  
only retaining the ``bubble'' diagram for $\Sigma(G)$, 
see Refs.~\cite{leticia,ABL,kunidave} for detailed discussions, 
and \cite{II} in the present context. 

\subsection{Three-point correlation: dynamic susceptibility 
and hydrodynamic contributions}
\label{sectiongiulio3pt}

From the above general inversion formulas for block matrices,  
Eq.~(\ref{matrixinversion}), one can obtain an expression of the
inverse of $\partial^2 \Gamma$ in the  ``$G\Psi$-sector'' in a form
transparent both from physical and diagrammatic standpoints. The off-diagonal 
block-element of Eq.~(\ref{matrixinversion}) gives, in particular, the energy-dynamics 
correlator (see Eq. (\ref{fdt})) and can be rewritten exactly as:
\be
S_T \equiv (\partial^2 \Gamma)^{-1,G\Psi} = -A^{-1}B \langle \Psi \Psi \rangle,
\ee
where we have used that $\langle \Psi \Psi \rangle \equiv (\partial^2 \Gamma)^{-1,\Psi\Psi}$.
Now, the equation determining the two-point correlators is   
$\delta \Gamma/\delta G=0$. Therefore the variation of the value 
of $G$ due to a small variation of $\Psi$, all other parameters being 
kept fixed,  
is given by: 
\be \label{chiTdiag} 
\frac{\delta G}{\delta \Psi} \delta \Psi 
=-\left[\frac{\delta^2 \Gamma}{\delta G \delta G}\right]^{-1}  
\frac{\delta^2 \Gamma}{\delta G \delta \Psi} \delta \Psi \equiv -A^{-1}B\,
\delta \Psi, 
\ee  
showing that the operator $\chi_\Psi= - A^{-1}B$ is the response of 
two-point correlators to a change in conserved quantities. 
Gathering these results, three-point functions
read~\footnote{The following identity
can also be seen as a Novikov formula for Gaussian fluctuations.}:
\be
\label{3-pointFT}
S_T = \chi_\Psi \langle \Psi \Psi \rangle,
\ee
providing an {\it exact} decomposition with a simple physical meaning. 
The correlation between energy at one point in space-time and dynamics 
elsewhere is {\it governed} by the 
sensitivity of the dynamics to energy changes, 
as encapsulated by $\chi_\Psi$, which contains all genuine collective 
effects in the dynamics induced by interactions.
This correlation is {\it mediated} 
by energy transport, $\langle \Psi \Psi \rangle$, 
with has a trivial hydrodynamic structure.

\begin{figure} 
\psfig{file=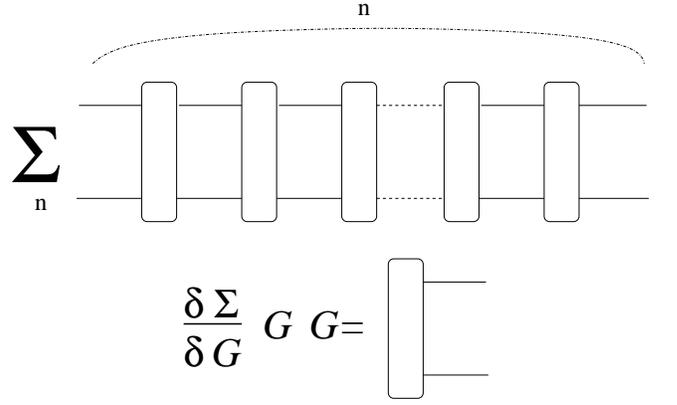,width=8.5cm} 
\caption{\label{ladders} 
Diagrammatic representation of the parquet diagrams 
obtained 
by expanding Eq.~(\ref{lad-eq}): $(1-\partial_G\Sigma \, GG)^{-1}
=\sum_n (\partial_G\Sigma \, GG)^n$.} 
\end{figure} 
 
In order to see this more clearly, let us now explore the diagrammatic content of $\chi_\Psi=A^{-1} B$. 
The three-leg vertex contribution $B=\frac{\delta^2 \Gamma}{\delta G \delta \Psi}$ is generically expected to be
non singular. The $A^{-1}$ term, on the other hand, can be rewritten using the general expression of $\Gamma$ as:  
\[ A^{-1} = \left[ 
\frac{\delta^2 \Gamma}{\delta G \delta G}\right]^{-1}=\left[G^{-1} 
G^{-1} -\frac{\delta 
    \Sigma(G)}{\delta G}\right]^{-1},   
\] 
where the objects in the above expression are four-index 
matrices. This term can be rearranged as follows:  
\ba \label{lad-eq}&& \left[ 
\frac{\delta^2 \Gamma}{\delta G \delta G}\right]^{-1}_{a,b;c,d} \equiv 
\sum_{c',d'}G_{a,c'}G_{b,d'} \nonumber \\ && 
\times 
\left( 
\delta_{c',c}\delta_{d',d}-\sum_{c'',d''}\frac{\delta 
\Sigma_{c'd'}(G)}{\delta G_{c'',d''}}G_{c'',c}G_{d'',d} \right)^{-1}. 
\ea  
One can now formally expand the term in parenthesis as 
$(1-\partial_G\Sigma \, GG)^{-1}=\sum_n (\partial_G\Sigma \, GG)^n$ to 
recover the so-called ``parquet'' diagrams~\cite{BlaizotRipka} that 
give a formally exact representation of the four-point function, see Fig.~\ref{ladders}.
This infinite series can provide a divergent contribution 
(as is the case within MCT  
at the critical point \cite{BB}), signaling the existence of 
a growing dynamical 
correlation length and non-trivial collective effects.

The important conclusion of this section is that the three-point 
function contains 
both a long-ranged hydrodynamical contribution $\langle \Psi \Psi \rangle$ 
related to energy conservation, and 
an interaction specific contribution -- the dynamical susceptibility 
$\chi_\Psi$. When specializing to the
integral over space of the three point function, as in Eq. (\ref{fdt}), 
the contribution of $\langle \Psi \Psi \rangle$ 
factors out and gives thermodynamic prefactors.
This shows that $\partial C/\partial T$ gives in fact a direct 
access to the dynamical susceptibility $\chi_\Psi$ at $q=0$.
Therefore, the lengthscale extracted from $\partial C/\partial T$ 
reveals the existence of collective dynamics, and is not 
related to any thermal diffusion or other hydrodynamical length.

\subsection{Four-point correlation functions and ensemble dependence} 

Let us now turn to a similar analysis of the four-point correlations.  
We start again from the above general inversion formulas for block matrices,  
Eqs.~(\ref{matrixinversion}, \ref{magicrelation}). In the simple case where the
$\Psi_a$'s are identically zero by symmetry, as  
happens for instance in the $p$-spin model for which a gauge symmetry
implies that the average value of the spins is   
always zero, the block matrix $B$ is also zero.   
Equation~(\ref{matrixinversion}) then 
simplifies to:
\be 
(\partial^2 \Gamma)^{-1,G}_{a,b;c,d} 
= \left[\frac{\delta^2 \Gamma}{\delta G_{a,b}\delta G_{c,d}}\right]^{-1} 
= A^{-1}. 
\ee 
In general this symmetry does not hold, in particular for liquids for
which the analysis is more  
involved. However, it turns out that $A^{-1}$ remains the fundamental object.
Using Eq.~(\ref{magicrelation})  
and the bottom right part of the matrix inversion relation,  
Eq.~(\ref{matrixinversion}), the four-point correlation functions 
can be written in a physically transparent way: 
\ba 
\label{gamma''} 
(\partial^2 \Gamma)^{-1,G}_{a,b;c,d} & = & 
\left[\frac{\delta^2 \Gamma}{\delta G_{a,b}\delta G_{c,d}}\right]^{-1} 
+ \nonumber \\ 
&& \sum_{e,f} \left(\frac{\delta G_{ab}}{\delta \Psi_e} \langle 
    \Psi_e \Psi_f \rangle_c 
\left(\frac{\delta G_{cd}}{\delta \Psi_f}\right)^{\dagger} \right). 
\ea 
This expression parallels Eq.~(\ref{landau}) in
Sec.~\ref{sectionjp}, and 
the last term corresponds to the dynamic fluctuations  
induced by the fluctuations of conserved quantities.
This formula is however much more general because it applies  
not only to $\chi_4(t)$ but also to $S_4(q,t)$. Indeed, in Fourier
space, the terms contributing to $S_4(q,t)$ read: 
\be 
\langle \delta \rho_{-k_3} (t) \delta \rho_{k_3+q}(0)  
\delta \rho_{-k_4} (t) \delta \rho_{k_4-q}(0)\rangle. 
\ee 
Therefore the extra contribution from conserved quantities, namely the
last term in Eq.~(\ref{gamma''}), reads: 
\begin{widetext} 
\begin{equation}\label{termchiT} 
\sum_{e,f}\int d\omega 
\frac{\partial \langle \delta \rho_{-k_3} (t) 
\delta \rho_{k_3+q}(0)\rangle}{\partial \Psi_{e}(\omega,q)} \langle 
    \Psi_{e}(\omega,q) \Psi_{f}(-\omega,-q)\rangle_c 
\frac{\partial \langle \delta \rho_{-k_4} (t) \delta \rho_{k_4-q}(0)\rangle } 
    {\partial \Psi_{f}(-\omega,-q)}. 
\end{equation}
\end{widetext} 
Now, one should notice that all terms corresponding to indices in the
response field sector of $\Psi$ (i.e. $e,f>m$)       
identically vanish at $q=0$. The reason is that the response fields
always appear in the vertices of the field-theory in the form 
$\nabla \Psi$. As a consequence, terms like
${\delta^2 \Gamma}/{\delta G \delta \Psi_{e}(\omega,q)}$ 
, for $e > m$, are proportional to $q$ at small $q$. 
 
In the case $q=0$, the value of conserved fields such as
$\Psi_{e}(\omega,q)$ for $e\leq m$ are by   
definition constant over time and set by initial conditions:   
$\langle \Psi_{e}(\omega,q) \Psi_{f}(-\omega,-q)\rangle_c=V
\delta(\omega) \Sigma_{ef}$ where   
$\Sigma_{ef}$ are the correlators of thermodynamic fluctuations of all
conserved quantities $\Psi$,   
determined by the probability distribution of initial conditions.  
As a consequence the term in Eq.~(\ref{termchiT}) at $q=0$ 
precisely reduces to the form discussed in the previous 
section on general grounds for $\chi_4(t) \equiv S_4(q=0,t)$: 
\begin{equation}\label{termchiTfin} 
\sum_{e,f=1}^{m} 
\frac{\partial \langle \delta \rho_{-k_3} (t) 
\delta \rho_{k_3}(0)\rangle}{\partial \Psi_e} \Sigma_{ef}  
\frac{\partial \langle \delta \rho_{-k_4} (t) \delta \rho_{k_4}(0)
\rangle }{\partial \Psi_f}. 
\end{equation} 
For Brownian dynamics, density is the only conserved quantity
and thus only one term, $m=1$, contributes 
to the sum in Eq.~(\ref{termchiTfin}).  
In the case of Newtonian dynamics there  
are in principle $2+d$ conserved quantities, density, momentum and
energy. However, by symmetry, the  
contribution of the momentum fluctuations is zero, and 
only density and energy should be considered. 
In $p$-spin disordered systems,   
on the other hand, this extra term is absent, and $m=0$.   
 
The conclusion is once again that the choice of statistical ensemble
matters for determining fluctuations of intensive dynamical correlators 
which correspond to $q=0$.  
For $q \neq 0$, the extra terms Eq.~(\ref{termchiT}) are in   
general always non-zero and contribute to $S_4(q,t)$.  
On the other hand, 
if one focuses on the case where $q=0$ exactly, 
the initial distribution is crucial.
As an example, in the case of Newtonian dynamics in the $NVE$ ensemble,  
all the extra contributions vanish since  
in that ensemble all conserved quantities are  
strictly fixed and $\Sigma_{ef}^{NVE} \equiv 0$.    
Thus, we find again within this formalism that 
the equality  
$\lim_{q\rightarrow 0}S_4(q,t)=S_4(0,t)$ is valid only in the 
ensemble where all conserved quantities fluctuates. In other
ensembles, such as $NVE$, the limit is singular.
  
\begin{figure} 
\psfig{file=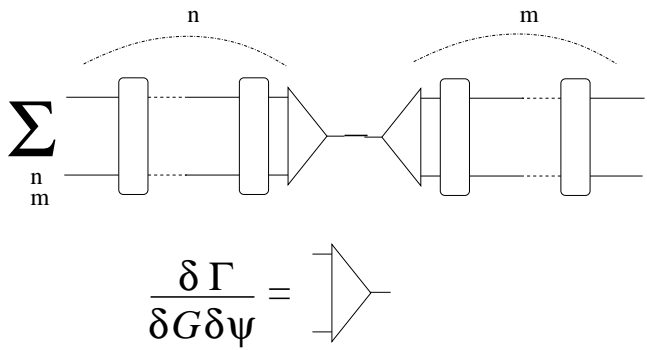,width=8.5cm} 
\caption{\label{figchit} 
``Squared-parquets'' representation of the contribution of
conserved quantity fluctuations to $\chi_4$. 
They correspond to the second term in Eq.~(\ref{gamma''}). 
} 
\end{figure}

Let us now explore the diagrammatic content of Eq.~(\ref{gamma''}). 
The first
term was already discussed in the previous section and can be 
expressed as a sum
of parquet diagrams, see Fig.~\ref{ladders}. The second term in 
Eq.~(\ref{gamma''}) also has a direct diagrammatic 
interpretation, shown in Fig.~\ref{figchit}. It consists of two 
parquets closed by three-leg vertices and joined by a correlation 
function of conserved variables. The wavevector $q$ in $S_4(q,t)$ is, in 
the diagrams, the wavevector flowing into the parquets and in the middle 
``link'' corresponding to $\langle \Psi_e \Psi_f \rangle_c$, as it 
appears explicitly in Eq.~(\ref{termchiT}). A detailed analysis of the 
structure of the diagrams shows that hydrodynamic scaling between time
and length is present only in the middle 
``link'' corresponding to $\langle \Psi_e \Psi_f \rangle_c$. 

As discussed in \cite{BB}, MCT provides a simple approximation in which the self-energy $\Sigma$ 
is approximated by the ``bubble'' diagram (see \cite{II}). Within this approximation
the parquet diagrams simplify into the ladder diagrams analyzed in \cite{BB}, which
diverge at the mode-coupling critical point. In Ref.~\cite{BB}, however, only the ladder diagrams were analyzed. The
contribution to $S_4$ of Eq.~(\ref{termchiT}) corresponding to the ``squared ladders'' was overlooked. 
As a consequence the MCT results in Ref.~\cite{BB} for $\chi_4(t)= S_4(q=0,t)$ only apply close to the critical point in the 
following cases (see also \cite{II} for further discussion):  
\begin{itemize} 
\item $NVE$ ensemble for Newtonian dynamics; 
\item $NVT$ ensemble for Brownian dynamics; 
\item $p$-spin models. 
\end{itemize} 
On the other hand, whenever conserved quantities are allowed to fluctuate, 
or when considering $S_4(q,t)$ at non-zero values of $q$, the contribution of (\ref{termchiT})  
may be important. For example, within the  context of MCT where $\chi_\Psi$  
diverges as $\epsilon^{-1}$ (where $\epsilon$ is the reduced distance from the 
critical point), the contribution of (\ref{termchiTfin}) in fact becomes
dominant and $\chi_4$ for Newtonian dynamics diverges much faster, as $\epsilon^{-2}$. However, see II for a 
discussion of the application of these MCT results
to real systems, where the MCT transition is avoided.

\subsection{A direct measure of dynamical susceptibility}

The analysis of the above sections show that in general $S_4(q,t)$ and $\chi_4(t)$ 
receive contributions of different physical origin with possibly 
different temperature dependencies, and whose relative amplitude 
might even depend on the chosen microscopic dynamics (Brownian or  
Newtonian). On the other hand, we have seen that all the interesting
physics is contained in the fundamental operator $A=[\delta^2 
\Gamma/\delta G \delta G]$, which governs the growth of dynamic   
correlations. Therefore it is both of theoretical and practical importance  
to introduce an observable with a physical content similar to that of $S_4(q,t)$, but
unaffected by the presence of global conservation laws and therefore by the choice
of statistical ensemble. Such an observable was discussed recently~\cite{BBMR}. It corresponds to the response of the 
intermediate scattering function (the two-point correlator $G$) to a  
small inhomogeneous external potential $V_{ext}$. Within the previous formalism one writes:  
\[ 
\frac{\delta^2 \Gamma}{\delta G \delta G}  
\frac{\delta G}{\delta V_{ext}}+\frac{\delta^2 \Gamma}{\delta G \delta 
    V_{ext}}+\frac{\delta^2 \Gamma}{\delta G \delta 
    \Psi}\frac{\delta \Psi}{\delta V_{ext}}=0 ,
\] 
and therefore: 
\[ 
 \frac{\delta G}{\delta V_{ext}}=-A^{-1}\left( 
\frac{\delta^2 \Gamma}{\delta G \delta 
    V_{ext}}+\frac{\delta^2 \Gamma}{\delta G \delta 
    \Psi}\frac{\delta \Psi}{\delta V_{ext}} 
\right).
\] 
Since the source term on which the operator $A^{-1}$ acts is expected to be only weakly temperature  
and density dependent, one sees that this quantity gives an almost direct  
measure of the critical behavior of the dynamic correlations encoded
in the operator $A$. When the external potential is homogeneous in space one
finds a quantity proportional to $\chi_\Psi$ above \cite{BBMR}, while for an
inhomogeneous external potential one can probe the full spatial structure of dynamic fluctuations. 
Indeed, when one differentiates with respect to the Fourier component $V_{ext}(q)$, the wavevector $q$ plays the same role
as for $S_4$~\cite{BBMR}. This can be seen at the diagrammatic level because $q$ is the 
wavevector entering into the ladders in Fig.~\ref{figchit}. 
  
\section{Physical consequences and issues} 
\label{sectiongiulioend} 

At this stage, it is important to summarize the conclusions drawn from the rather dense theoretical
analysis presented above. This will allow us to identify clearly the questions that need to be tested numerically 
before possibly extrapolating these conclusions to real glass-forming systems. 

We established in the previous section that all non-trivial collective dynamical effects are encoded into a certain operator $A^{-1}$, 
which could in principle be reached by measuring the sensitivity of the local dynamics to an external potential~\cite{BBMR}. 
More easily accessible quantities are derivatives of two-point correlations with respect to temperature, or density. 
We have shown in detail how these are indeed proportional to $A^{-1}$
and provide lower bounds on $\chi_4$, and are therefore of direct interest to probe the
growth of a dynamic lengthscale in glasses, as claimed in~\cite{science}. 
However, the assumption that growing susceptibilities imply growing 
lengthscales needs to be discussed more thoroughly.

\subsection{Growing susceptibilities vs growing lengthscales}
 
The first important remark is that the lower bound on $\chi_4$ obtained 
in the previous sections is useful only when $\chi_4$ is significantly larger than one because
$\chi_4$ is of the order one even in an ideal gas \cite{TWBBB}.
The second remark is that one has to be sure that the growth of the 
susceptibility is due to a growing length and not to growing 
local fluctuations. For simplicity,
suppose that the energy-dynamics correlator $S_T(y,\tau_\alpha)$ can be 
written as $S_T(0) \times y^{2-d-\eta} f(y/\xi)$. Its space integral $\chi_T$ is then given by:
\be
\chi_T = S_T(0) \xi^{2-\eta} \int d^d u u^{2-d-\eta} f(u).
\ee
This shows the origin of an increase in $\chi_T=\partial C/\partial T$ as $T$ decreases is due to 
\begin{itemize}
\item {either} an increase in $\xi$ with a roughly constant $S_T(0)$
\item {or} because $S_T(0)$ increases whereas $\xi$ is trivial,
\end{itemize}
or, of course, through a combination of both. In order to be confident that the first scenario is 
the correct one, and that $\chi_T$ can be used to estimate a correlation volume, one needs to be sure that $S_T(0)$ is of order one
and basically temperature independent. This requires in principle some extra information, for example on the full spatial 
dependence of $S_T(y,\tau_\alpha)$. This will be checked in numerical simulations below. We note also
that MCT precisely realizes the first scenario above.  

From a physical point of view, one expects the enthalpy fluctuations 
$\delta h$ to contain a fast (kinetic) part and a slow (configurational) part, of
similar order of magnitude ($k_B T$). While it is clear that the fast part should have very small correlations with the local 
correlation on time scale $\tau_\alpha$, there is no reason to think that $\langle o(x,\tau_\alpha)o(x,t=0) \delta h_{slow}(x,t=0) 
\rangle$ is particularly small. Quite the contrary, we expect that this is of order $k_B T$ in glassy systems. But interestingly,
this suggests that the specific heat $c_P$ that should enter the relation between $\chi_T$ and $\xi$ should be the so-called
excess specific heat $\Delta c_P$, restricted to slow (glassy) degrees of freedom, as surmised in \cite{science}.

\subsection{Statistical ensemble and dynamics 
dependence of dynamic fluctuations}

A rather bizarre 
conclusion of the previous section is that global 
four-point correlators, corresponding to the fluctuations 
of intensive dynamical
correlators, not only depend on the   
{\it statistical ensemble} (for $q=0$) but, remarkably and perhaps unexpectedly, also on the {\it choice of dynamics}.  
This is to be contrasted with the case for two-point correlators, which are independent of the chosen ensemble and are known 
numerically to be independent of the dynamics, at least in the relevant ``slow'' regime~\cite{gleim,szamel2,tobepisa}. 
This shows that four-point correlators, although containing 
some useful information on dynamical heterogeneities, 
mixes it with other, less interesting physical effects. 
Clear-cut statements with four-point quantities  
can however be made when the dynamic lengthscale grows substantially at some finite temperature or density,  
as for example within MCT where the operator $A^{-1}$  develops a zero mode that leads to a divergence of the dynamic lengthscale 
$\xi$. When the dynamic lengthscale becomes very large, these statements may be summarized as follows~:\begin{itemize} 
\item $\chi_4(\tau_\alpha)$ for $NVT$ Newtonian dynamics 
diverges more strongly than $\chi_4(\tau_\alpha)$ for $NVT$ 
stochastic dynamics; 
\item $\chi_4(\tau_\alpha)$ for $NVE$ Newtonian dynamics 
diverges like $\chi_4(\tau_\alpha)$ for $NVT$ stochastic dynamics; 
\item $\chi_4(\tau_\alpha)$ for $NVE$ Newtonian dynamics and $NVT$
      stochastic dynamics diverge like $\chi_T(\tau_\alpha)$  
(or $\chi_\rho(\tau_\alpha)$). 
\end{itemize} 
We will test these statements numerically in the next section, and will indeed
establish that $\chi_T$, $\chi_4^{NVE}$ and $\chi_4^B$ increase in exactly the same way
with $\tau_\alpha$.
The full time dependence of these  
different correlators will be discussed  
in the companion paper, Ref.~\cite{II}.

\subsection{A unique dynamic correlation length}
 
Let us emphasize again that although $\chi_4(\tau_\alpha)$ for $NVT$ Newtonian 
dynamics and stochastic dynamics diverge differently, our results strongly suggests these quantities in fact reflect the same 
underlying physics, which is the growth of a {\it unique} lengthscale 
$\xi$ in all of these cases. Only the relation between 
$\chi_4(\tau_\alpha)$ or $\chi_T(\tau_\alpha)$ and $\xi$ changes: dynamic
fluctuations are amplified because of conserved variables.
This becomes clear when one considers the (ensemble-independent) function 
$S_4(q,\tau_\alpha)$ for $q \neq 0$.  In all of these cases, $S_4(q,\tau_\alpha)$ can be 
written as a scaling function $g_4(q \xi)$ with the same $\xi$ but 
different functional forms. For example, $g_4^{N}(q \xi)$ for 
Newtonian dynamics can be written as $g_4^{B}(q \xi)+ c_q [g_4^{B}(q 
\xi)]^2$, where $c_q$ is a coefficient and $g_4^{B}(q \xi) \sim g_T(q \xi)$ the 
scaling function for Brownian dynamics or governing $S_T(q,\tau_\alpha)$. 
Note that the relation between $\chi_4(\tau_\alpha)$ for $NVE$ Newtonian dynamics, 
$\chi_4(\tau_\alpha)$ for $NVT$ Brownian dynamics and 
$\chi_T(\tau_\alpha)$ may not be accurate far from any critical point, since these
quantities are affected by different, non-critical prefactors. 

\subsection{Response vs correlation functions}
 
Four point correlators were originally hoped to be suited to 
quantify precisely dynamical heterogeneities in glass-formers, 
as motivated in Sec. \ref{sectionjp}. The conclusion of the previous section 
and the numerical results of the following ones show that although they contain
indeed crucial information, it is mixed up with less interesting 
physical effects. Nevertheless, a unique dynamic correlation length seems
to govern the slowing down independently of the dynamics, global dynamic
fluctuations depend on the dynamics and on the ensemble.
As discussed formally in the previous section, response functions measuring the
response of the dynamics to local perturbations do not present these 
difficulties. They should be independent of the microscopic dynamics, as it is 
the case for two-point correlators, and probe directly the dynamic 
correlations without mixing them up with other effects due to conservation 
laws.

In the following, we give some numerical evidence for the most important claims made in this 
paper: the existence of unique lengthscale $\xi$ governing the growth of $\chi_4$ and $\chi_T$ 
and the ensemble and dynamics dependence of the four-point correlators.
 
\section{Numerical results for two molecular glass-formers} 
 
\label{MD} 

We now present our numerical calculations of the dynamic 
susceptibility $\chi_T(t)$, its relation to $\chi_4(t)$, and the 
behavior of spatial correlations $S_T$ and $S_4$ in two well-studied models of 
molecular glass-formers: a binary Lennard-Jones (LJ) 
mixture~\cite{KA}, considered as a simple model system for fragile 
supercooled liquids~\cite{hans}, and the Beest, Kramer, and van Santen 
(BKS) model, which is a simple description of the strong glass-former 
silica~\cite{beest90,bks_sim}.  A first motivation for these 
simulations is that all terms contributing to the dynamic fluctuations 
can be separately evaluated and quantitatively compared. 
Spatial correlators and dynamic 
lengthscales can be directly 
evaluated in the simulations to confirm the link between 
dynamic susceptibilities and dynamical lengthscales. Therefore, 
the claim made in Ref.~\cite{science} that $\chi_T(t)$ yields direct 
experimental access to a dynamical lengthscale can be 
quantitatively established.  A second interesting feature is that the 
influence of the microscopic dynamics and statistical ensemble can be 
quantified in the simulations by keeping the pair potential unchanged 
but by switching from the energy conserving Newtonian dynamics to some 
stochastic dynamics which locally supplies energy to the particles.

\subsection{Models and technical details} 
 
The binary LJ system simulated in this work is a 80:20 mixture 
of $N_A=800$ and $N_B=200$  
Lennard-Jones particles of types $A$ and $B$, with interactions 
\begin{equation} 
\phi_{\alpha \beta}^{\rm LJ}(r)= 
4 \epsilon_{\alpha \beta} \left[ 
\left(\frac{\sigma_{\alpha \beta}} {r} 
\right)^{12} - 
\left(\frac{\sigma_{\alpha \beta}}{r} 
\right)^{6} 
\right], 
\end{equation} 
where $\alpha, \beta \in [{A}, {B}]$ and 
$r$ is the distance between the particles of type $\alpha$ and $\beta$. 
Interaction parameters $\epsilon_{\alpha \beta}$
and $\sigma_{\alpha \beta}$
are chosen to prevent crystallization
and can be found in Ref.~\cite{KA}. 
The length, energy and time units are the standard 
Lennard-Jones units $\sigma_{AA}$ (particle diameter), 
$\epsilon_{AA}$ (interaction energy), and $\tau_0 = 
\sqrt{m_A\sigma_{AA}^2/ (48 \epsilon_{AA})}$, where $m_A=m_B$ 
is the particle mass and the subscript $A$ refers to the majority 
species. Equilibrium 
properties of the system have been fully 
characterized~\cite{KA}. 
At the reduced density $\rho_0=1.2$, where all our 
simulations are carried out, the MCT transition has been  
conjectured to be in the vicinity of $T_c \approx 0.435$ \cite{KA}. 
The slowing down of the 
dynamics, $T \gtrsim 0.47$, can be correctly described by mode-coupling 
theory, but this description eventually  breaks down  
when lowering the temperature further, 
$T \lesssim 0.47$~\cite{KA}. 
 
To check the generality of our results, we have also investigated  
the behavior of a second glass-former, characterized by a very different 
fragility. To this end, we simulate   
a material with an Arrhenius dependence  
of its relaxation time, namely silica. 
Various simulations have shown that a reliable pair potential  
to simulate silica is the one 
proposed by BKS~\cite{beest90,bks_sim}. 
The functional form of the BKS potential is 
\begin{equation} 
\phi_{\alpha \beta}^{\rm BKS}(r)= 
\frac{q_{\alpha} q_{\beta} e^2}{r} + 
A_{\alpha \beta} \exp\left(-B_{\alpha \beta}r\right) - 
\frac{C_{\alpha \beta}}{r^6}, 
\label{eq1} 
\end{equation} 
where $\alpha, \beta \in [{\rm Si}, {\rm O}]$ and $r$ is the distance 
between the ions of type $\alpha$ and $\beta$.  The values of the 
constants $q_{\alpha}, q_{\beta}, A_{\alpha \beta}, B_{\alpha \beta}$, 
and $C_{\alpha \beta}$ can be found in Ref.~\cite{beest90}. For the 
sake of computational efficiency the short range part of the potential 
was truncated and shifted at 5.5~\AA. This truncation also has the 
benefit of improving the agreement between simulation and experiment 
with regard to the density of the amorphous glass at low temperatures. 
The system investigated has $N_{\rm Si} = 336$ and $N_{\rm O}=672$ 
ions in a cubic box with fixed size $L=24.23$~\AA.  The Coulombic part 
of the potential has been evaluated by means of the Ewald sum using a 
constant $\alpha L$=10.177. 
 
For both LJ and BKS models we have numerically 
integrated Newton's equations of 
motion using the velocity Verlet algorithm~\cite{at} using a time-step  
$h^{\rm LJ} =0.01 \tau_0$  
and $h^{\rm BKS}=1.6$~fs, respectively. Doing so we can 
measure  
spontaneous dynamic fluctuations in the microcanonical 
$NVE$ ensemble. Before these microcanonical production runs,  
all systems are equilibrated using a stochastic  
heat bath for a duration significantly longer than the typical  
relaxation time, $\tau_\alpha$, implying that particles  
move over several times their own diameter during equilibration.    
Production runs were at least larger than $30\tau_\alpha$, 
and statistical convergence for dynamic fluctuations  
was further improved by simulating 10 independent samples of each 
system at each temperature.  
Repeating this strategy for many temperatures in two molecular systems  
obviously represents a substantial numerical effort. 
 
To check the influence of the microscopic 
dynamics, and in particular the role of the  
energy conservation, we 
have also performed stochastic simulations of the LJ system  
using two different techniques. Following Ref.~\cite{gleim}, 
we have simulated Brownian dynamics where Newton's equations 
are supplemented by a random force and a viscous friction 
whose amplitudes are related by the fluctuation-dissipation  
theorem. The numerical algorithm used to integrate these 
Brownian equations of motion is described in Refs.~\cite{at,gleim} 
using the time step of $h^{\rm LJ}_{\rm BD}=0.016 \tau_0$ 
and a friction coefficient $\zeta = 10 m\tau_0$. 
We use the equilibrium configurations obtained by MD simulations
as starting point for our production runs in Brownian simulations. 
Finally we have implemented a second stochastic dynamics,  
a standard Monte Carlo dynamics,  
with the LJ potential~\cite{tobepisa}.  
At time $t$, the particle $i$, located at the position ${\bf r}_i(t)$,  
is chosen at random. The energy cost $\Delta E$ to move it to the new 
position ${\bf r}_i(t) + {\boldsymbol \delta}$ is evaluated,   
${\boldsymbol \delta}$ being a random vector comprised in a square of 
lateral size  
$\delta_{\rm max} = 0.15$. The Metropolis acceptance rate, $p = {\rm min} 
(1, e^{-\beta \Delta E})$, is then  
used to decide whether the move is accepted~\cite{at}.  
One Monte Carlo time-step 
represents $N=N_A + N_B$ attempts to make such a move. 

For BKS, we only present results for ND  
because BD simulations at low enough temperature would be numerically
too costly in this system. The reason is that
a very large friction coefficient is needed 
to have a truly damped dynamics~\cite{unpub}, 
making the overall relaxation much too slow 
to be studied numerically at low temperature.
Monte Carlo simulations are similarly slow because of the 
long-range character of the Coulomb term in the BKS potential.
Very recently we have developed a short-range 
approximation of the BKS potential that allows
much faster Monte-Carlo simulations, and we shall mention 
some preliminary results obtained for dynamic 
susceptibilities using this method~\cite{ludounpub}.  

\subsection{Physical observables} 
 
Following previous work~\cite{steve,berthier,TWBBB},  
we monitor the dynamical behavior  
of the molecular liquids through the self-intermediate 
scattering function, 
\be F_s({\bf k},t) = 
\left\langle \frac{1}{N_\alpha} \sum_{j=1}^{N_\alpha} e^{i {\bf k} 
\cdot [{\bf r}_j(t) - {\bf r}_j(0)]} \right\rangle, \label{self} 
\ee 
where the sum in Eq.~(\ref{self}) runs over one of the species of the 
considered liquid ($A$ or $B$ in the LJ, Si or O for silica). We 
denote by $f_s({\bf k}, t)$ the real part of the instantaneous value 
of this quantity, so that we have $F_s({\bf k}, t) = \langle 
f_s({\bf k},t)\rangle$. 
 
The four-point susceptibility, $\chi_4(t)$, quantifies the strength of 
the spontaneous fluctuations around the average dynamics by the 
variance, \be 
\chi_4(t) = N_\alpha \left[  \langle f_s^2({\bf k}, t) 
\rangle - F_s^2({\bf k}, t) \right]. \label{chi4lj} \ee In 
principle, $\chi_4(t)$ in Eq.~(\ref{chi4lj}) retains a dependence on 
the scattering vector ${\bf k}$.  
Since the system is isotropic,  
we circularly average (\ref{self}) and 
(\ref{chi4lj}) over wavevectors of fixed modulus. 
Note that the value of dynamical correlations depend on $|{\bf k}|$ 
as shown in \cite{lacevic,dauchot}. 
A detailed analysis of this dependence has been performed in 
\cite{Ck} and will be further discussed in
\cite{II}. In the following we will focus only on the value of  
$|{\bf k}|$ for which the dynamical correlations are more pronounced 
and that measure the correlation of the local dynamics. 
For the LJ system we will mainly consider results for $|{\bf k}| = 7.21$ 
and for the BKS one $|{\bf k}| = 1.7$~\AA$^{-1}$. 
These values respectively represent the typical distance 
between $A$ particles, and the size of the SiO$_4$ tetrahedra.
As discussed above, we expect $\chi_4(t)$ to depend on the chosen  
statistical ensemble, e.g. $NVE$ or $NVT$, for Newtonian dynamics,  
and to depend also on which microscopic dynamics is chosen,  
stochastic or energy conserving.   
 
To evaluate the temperature derivatives involved in \be \chi_T(t) = 
\frac{\partial}{\partial T} F_s({\bf k},t), \ee we perform 
simulations at nearby temperatures, $T$ and $T+\delta T$,   
and estimate $\chi_T(t)$ through 
finite differencing, $\chi_T(t) \approx \delta F_s(k,t) / \delta T$,  
as illustrated by arrows in Fig.~\ref{figBKS1} 
in Sec.~\ref{sectionjp}. For this procedure to be effective, temperature 
differences must be small enough that linear response holds.  
Taking $\delta T$ too small leads however to poor statistics.  
The smallest $\delta T$ which might be used can be estimated by comparing  
the statistical noise of $F_s(k,t)$ to the expected response  
$\chi_T \times \delta T$. This leads in our case to  
the typical lower bound $\delta T / T > 0.005$. We have  
typically used $\delta T / T \approx 0.01$, which is not far from  
the lower bound. For some selected temperatures, we have 
explicitly checked that linear response is satisfied by comparing  
results for $2\delta T$, $\delta T$ and $\delta T / 2$.  
 
It might be worth recalling that the value of $\chi_T(t)$ 
does not depend whether one works in $NVE$ or $NVT$, since  
ensemble equivalence obviously holds for this local 
observable~\footnote{Note, however, that in general one has to be careful 
that the derivative with respect to temperature
is performed keeping the same other variables fixed.}.    
Much less trivial is the numerical finding that $\chi_T(t)$  
is also found to be the same for Newtonian, Brownian and Monte Carlo  
dynamics for times pertaining to the structural  
relaxation. This directly follows from the 
non-trivial numerical observation that 
the average structural relaxation  
dynamics of the binary LJ system 
has no dependence upon its microscopic dynamics, 
apart from an overall time rescaling.  
On the other hand, the short-time dynamics is different  
in the three cases. 
Our findings then confirm for Brownian dynamics, and extend for  
Monte Carlo dynamics~\cite{tobepisa},  
the results of Refs.~\cite{gleim,szamel2} about the  
independence of the {\it average} glassy dynamics upon the microscopic  
dynamics. We will see below that clear differences emerge at the level of  
the {\it dynamic fluctuations}. 
 
\subsection{Amplitude of the dynamic fluctuations} 
 
In this paper, we restrict our analysis of the dynamic 
susceptibilities to the amplitude of 
the peaks observed in Fig.~\ref{figBKS1},  
meaning that we study dynamic fluctuations  
on a timescale $t \approx \tau_\alpha$. The time dependence of the  
fluctuations are studied in the companion paper, Ref.~\cite{II}. 

Furthermore, as discussed in Sec.~\ref{gf}, the contribution to $\chi_4^*$ 
due to density fluctuations in Eq.~(\ref{landau})
is significantly less than 
the one corresponding to energy fluctuations for most molecular liquids. 
Therefore, we will neglect the role of density fluctuations in 
the following
and focus only on $\chi_4^{NVT}$ since we expect that 
$\chi_4^{NPT}\simeq \chi_4^{NVT}$. As a more quantitative check,  
we have used the data in Ref.~\cite{sri} to 
estimate that the contribution of density
fluctuations to $\chi_4$ is about 10 times smaller than 
the temperature contribution for the LJ system 
at $\rho_0=1.2$. 
 
\begin{figure} 
\psfig{file=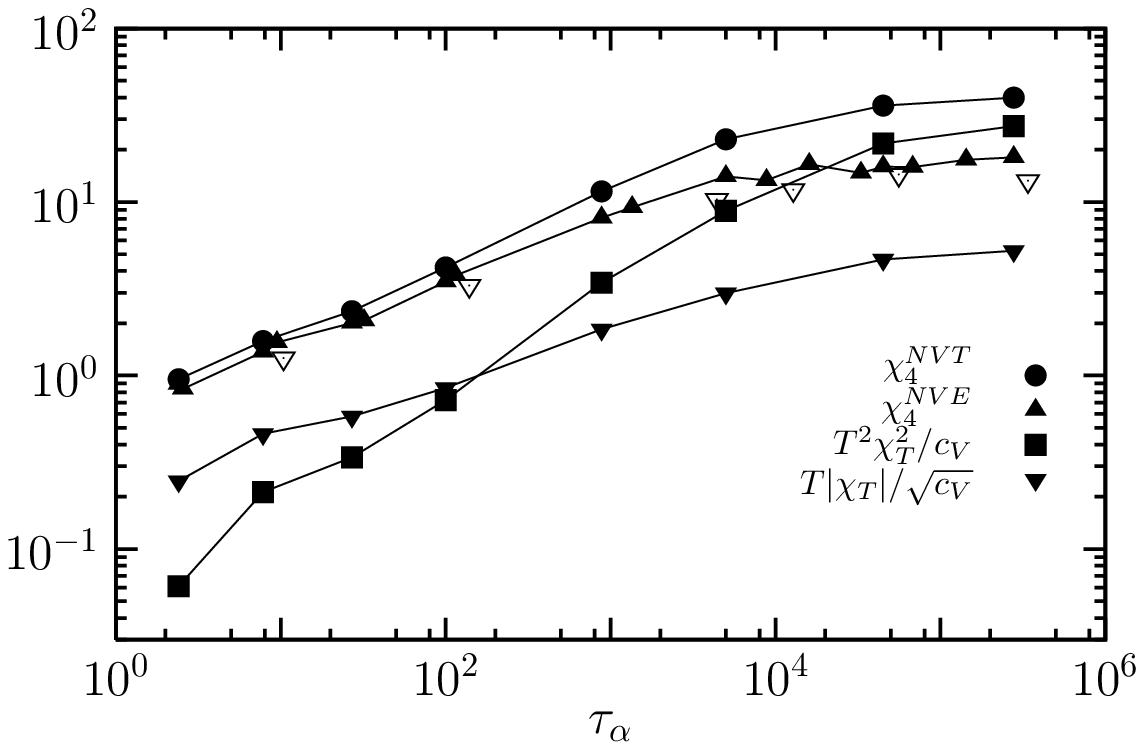,width=8.5cm} 
\psfig{file=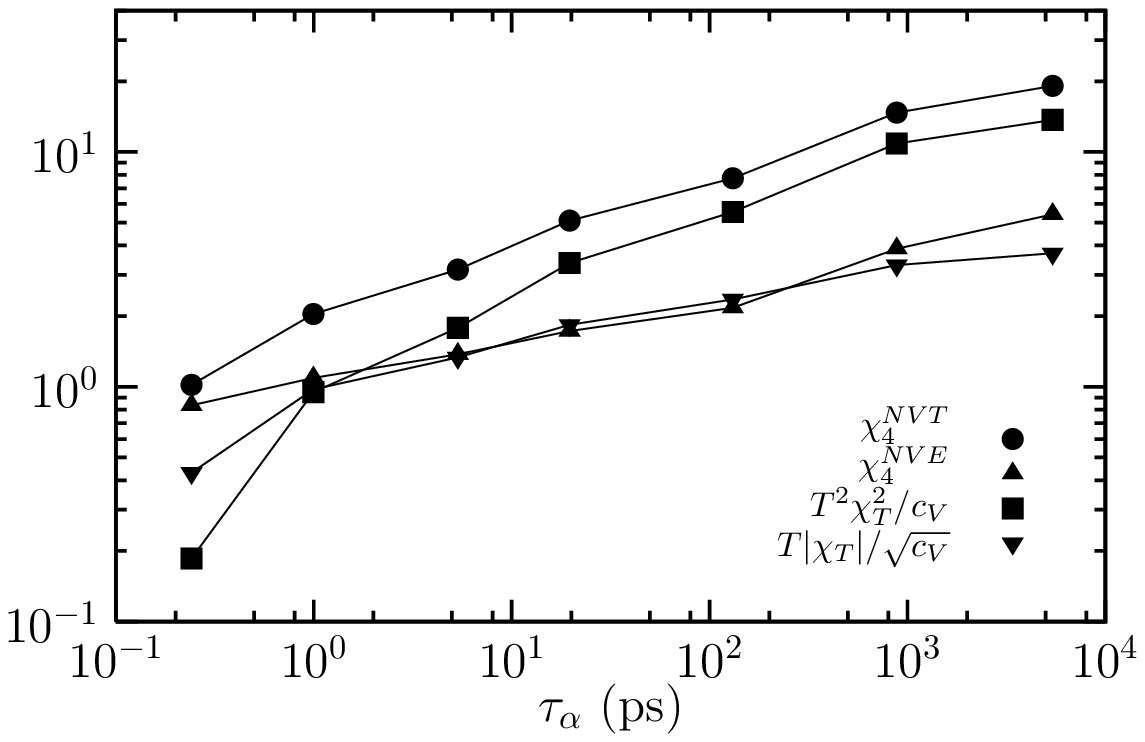,width=8.5cm} 
\caption{\label{figLJ3}  
Peak amplitude of various dynamic  
susceptibilities in the binary LJ mixture obtained  
from the $A$ particles dynamics (top) and  
the BKS model for silica from the Si ions dynamics (bottom). 
Open triangles in the LJ system represent $\chi_4^{NVE}$ measured  
in a smaller system with $N=256$ instead of the $N=1000$ 
used everywhere else in the paper.  
In both cases, $T^2 \chi_T^2 /c_V$ is smaller than $\chi_4^{NVE}$  
at high temperature, but increases faster and becomes  
eventually the dominant contribution to $\chi_4^{NVT}$  
in the relevant low temperature glassy regime. Note that the crossing occurs
much earlier for BKS.} 
\end{figure} 

\subsubsection{Ensemble dependence of dynamical correlations}

Our results are summarized in Fig.~\ref{figLJ3}, where  
we present our numerical data for $T |\chi_T| /\sqrt{c_V}$,  
$\chi_4^{NVE}$, $T^2 \chi_T^2 /c_V$ and  
the sum $\chi_4^{NVT} = \chi_4^{NVE} + T^2 \chi_T^2 /c_V$,  
all quantities obtained from   
Newtonian dynamics simulations of both the LJ and BKS models. Recall that
we define $c_V$ in units of $k_B$ throughout the paper.
When temperature decreases, all peaks shift to larger times  
and track the $\alpha$-relaxation. Simultaneously, their height 
increases, revealing increasingly stronger dynamic correlations as 
the glass transition is approached.  
    
The main observation from the data displayed in Fig.~\ref{figLJ3}, 
already announced in Ref~\cite{science}, is that in both LJ and BKS 
systems the term $T^2 \chi_T^2 / c_V$ while being small, 
$\sim~O(10^{-1})$, above the onset temperature of slow dynamics, grows 
much faster than $\chi_4^{NVE}$ when the glassy regime is entered. As 
a consequence, there exists a temperature below which the temperature 
derivative contribution to the four-point susceptibility 
$\chi_4^{NVT}$ dominates over that of $\chi_4^{NVE}$, or is at least 
comparable.  This crossover is located at $T\approx 0.45$ in the LJ 
system, $T \approx 4500$~K for BKS silica.  The conclusion 
that $T^2 \chi_T^2 /c_V$ becomes larger than $\chi_4^{NVE}$ at low 
temperatures holds for both strong and fragile glass-formers, but 
for different reasons. In the LJ systems $\chi_T$ increases 
very fast because timescales grow in a super-Arrhenius manner, which 
makes the temperature derivative larger and larger, while 
$\chi_4^{NVE}$ saturates at low $T$.  In the BKS system, although the 
temperature derivative is not very large because of the simple 
Arrhenius growth of relaxation timescales, $\chi_4^{NVE}$ is even 
smaller~\cite{vogel}, i.e. much smaller than in the fragile LJ system.
It is interesting to note that the common value of $\chi_4^{NVE}$ and 
$T^2 \chi_T^2 /c_V$ when they cross is substantially larger for 
the LJ system ($\sim 10$) 
than for BKS ($\sim 1$). It would be interesting to see, more 
generally, how $\chi_4^{NVE}$ and
fragility are correlated.
 
It is important to remark that finite size effects  
could play a role in the present study:  
when measured in a system which is  
too small, dynamic fluctuations are underestimated~\cite{fss}.  
Therefore it could be that using too small a system we have  
underestimated $\chi_4^{NVE}$, and therefore  
observed a fictitious saturation of the inequality~(\ref{chi4bound}). 
To investigate  this possibility, we have included in 
Fig.~\ref{figLJ3} data for $\chi_4^{NVE}$ obtained  
in a system comprising  about 4 times less particles,  
$N=256$, with essentially similar results.  
We have checked that also the average dynamics is  
unchanged when $N=256$, so that $\chi_T$ is not  
affected by finite size effects either for the range of parameters
chosen. We are therefore  
confident that the main conclusion  
drawn from Fig.~\ref{figLJ3} is not an artifact due to finite size effects.  
 
We can therefore safely conclude that $T^2 \chi_T^2/c_V$ is an
excellent approximation to $\chi_4^{NVT}$ for relaxation times larger than  
$\tau_\alpha \approx 10^4$ in the LJ system,  
$\tau_\alpha \approx 10$~ps in BKS silica. Our results 
indicate that this becomes an even better approximation as temperature is  
lowered, at least in the numerically accessible regime. As reported in Ref.~\cite{science}, this suggests a direct 
experimental determination of $\chi_4$ close to the glass 
transition temperature, $T_g$.  
Our data indicate however, that care must be taken  
when analyzing the first few decades of the dynamical slowing down where  
all terms contribute differently to $\chi_4^{NVT}$, and have  
different temperature dependences~\cite{science,Ck,szamel}. 
We now show that despite their different temperature 
behavior, $\chi_4^{NVT}$, 
$\chi_4^{NVE}$
and $\chi_T$ contain the same physics, 
as predicted theoretically in previous sections.

\subsubsection{Dynamics dependence of dynamical correlations}
 
\begin{figure} 
\psfig{file=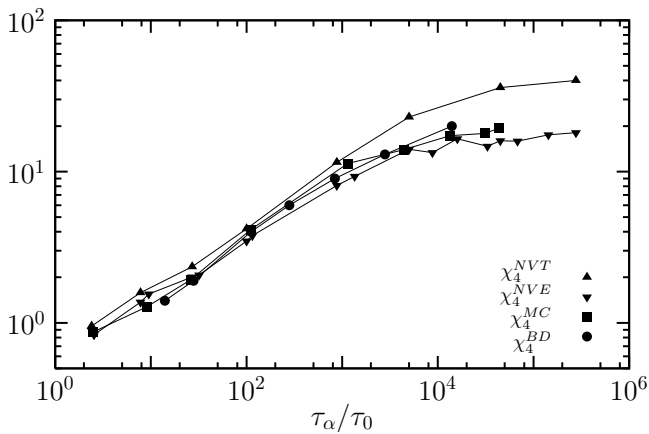,width=8.5cm} 
\caption{\label{figLJ4}  Amplitude of 
four-point susceptibilities $\chi_4(\tau_\alpha)$ obtained  
from the $A$ particles dynamics in the LJ system  
for Newtonian canonical ($\chi_4^{NVT}$), microcanonical  
($\chi_4^{NVE}$) dynamics and 
stochastic Monte Carlo ($\chi_4^{MC}$) 
and Brownian ($\chi_4^{BD}$) dynamics. 
Stochastic dynamics measurements follow the results obtained  
from microcanonical Newtonian dynamics, while the amplitudes
obtained in the canonical ensemble for Newtonian
dynamics are much larger, 
as predicted in Sec.~\ref{sectiongiulioft}.}  
\end{figure} 
 
We conclude this section with a discussion of  
the data for dynamic fluctuations obtained through our stochastic  
simulations. The temperature evolution of  
the dynamic susceptibilities $\chi_4(\tau_\alpha)$  
obtained with Monte Carlo and Brownian dynamics  
are shown in Fig.~\ref{figLJ4} where it 
is compared to the data obtained in both canonical and microcanonical  
ensembles with Newtonian dynamics. 
Our data unambiguously show that dynamic fluctuations  
with stochastic dynamics are different from the ones obtained  
with Newtonian dynamics in the $NVT$ ensemble. They are however  
very similar to the microcanonical ones.  
This result is not immediately intuitive 
because one could have imagined that stochastic simulations  
are a good approximation to the dynamics of liquids in the  
canonical ensemble. However, we have shown in  
Sec.~\ref{sectiongiulioft} that this naive expectation 
is in fact incorrect. The absence of the energy conservation  
in the stochastic dynamics (MD or BD) removes 
the contribution of the ``squared parquets'', which corresponds to the 
enhancement of dynamic fluctuations due to energy fluctuations,
and leads to $\chi_4^{NVE} \sim \chi_4^{BD} \sim \chi_4^{MC}$. 
This is in excellent agreement  with our numerical data. 

\begin{figure} 
\psfig{file=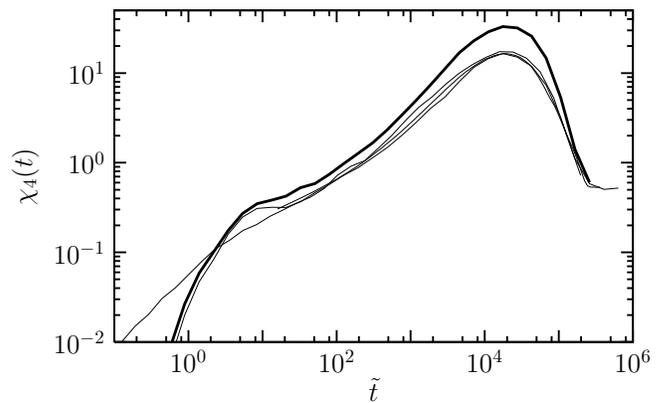,width=8.5cm} 
\caption{\label{giuliopenible}  
Four-point susceptibilities at $\chi_4(t)$ at $T=0.45$ 
obtained  from the $A$ particles dynamics in the LJ system  
for Newtonian canonical (shown as a thicker line), 
microcanonical dynamics and 
stochastic Monte Carlo and Brownian dynamics as a function of 
a rescaled time chosen so that all $\chi_4$'s overlap near the alpha relation.
We chose $\tilde{t}=t$ for $NVE$ Newtonian dynamics, $\tilde{t}=t/24$ 
for Brownian dynamics, $\tilde{t}=t/100$ for Monte Carlo dynamics.
The Newtonian $\chi_4^{NVT}(t)$ is larger than the others,
which are all nearly identical in both beta and alpha regimes.}
\end{figure}

Another confirmation of our theoretical expectations is 
presented in Fig.~\ref{giuliopenible} 
in which we show the time dependence of $\chi_4$
for $NVE$ Newtonian dynamics, $NVT$ Brownian, 
Monte-Carlo and Newtonian dynamics. 
The first three curves are essentially identical apart at 
microscopic times, whereas the last one is clearly larger.
This dependence on the microscopic dynamics 
is a general result obtained from the previous diagrammatic
discussion.

A further crucial prediction of our diagrammatic analysis is that 
$\chi_T$ and $\chi_4^{NVE}$ should have a similar critical scaling
 in temperature and time. This is again a general result 
if the three-leg vertex does not introduce any
additional singular behavior. In fact, as discussed in the 
previous sections, $\chi_T$ consists of a parquet diagram 
closed by a three-leg vertex whereas $\chi_4^{NVE}$ is given by single parquet 
diagrams. In Fig.~\ref{figLJ3} we confirm numerically
that the peaks of 
$\chi_T$ and $\chi_4^{NVE}$ scale in the same way with the temperature, 
both in the
LJ and BKS systems.
This similarity should in fact extend to the whole time dependence but the 
results are somewhat less satisfactory, as discussed in the companion 
paper~\cite{II}. 
   
For BKS we do not have numerical results 
for Brownian Dynamics for reasons mentioned above. 
However, our preliminary 
results from Monte Carlo simulations of a slightly modified 
version of the BKS potential~\cite{ludounpub} 
agree with the conclusions
drawn from the LJ data, that is, $\chi_4^{MC}$
seems to follow more closely  $\chi_4^{NVE}$, as in
Fig.~\ref{figLJ4},
with similar time dependences for the dynamic susceptibilities,
as in Fig.~\ref{giuliopenible}. 

\subsection{Spatial correlations} 

We now discuss the spatial correlations  
associated with the global fluctuations measured through 
$\chi_T(t)$ and $\chi_4(t)$. To this end, we 
define the local fluctuations of the dynamics  
through the spatial fluctuations of the instantaneous value 
of the self intermediate scattering function,  
\be 
\delta f_k({\bf x},t) =  \sum_i \delta( {\bf x} - {\bf r}_i(0))  
\left[ \cos [ {\bf k} \cdot (  
{\bf r}_i(t) - {\bf r}_i(0)) ] 
- F_s({\bf k},t) \right].  
\ee 
In the following, we will drop the ${\bf k}$ dependence  
of the dynamic structure factors to simplify notations. 
Local fluctuations of the energy at time $t$ 
are defined as usual,   
\be 
\delta e({\bf x},t) = \sum_i \delta( {\bf x} - {\bf r}_i(t))  
\left[ e_i(t) - e \right],  
\ee 
where $e_i(t) = m v_i^2(t)/2 + \sum_j V(r_{ij}(t))$ is the instantaneous  
value of the energy of particle $i$, and  
$e \equiv \langle N^{-1} \sum_i e_i \rangle$  is the average 
energy per particle. 
 
Spontaneous fluctuations of the dynamics can be detected  
through the ``four-point'' dynamic structure factor,  
\be  
S_4({\bf q},t) = \frac{1}{N} \langle  
\delta f({\bf q},t) \delta f(-{\bf q},t) \rangle, 
\label{s4} 
\ee 
while correlation between dynamics and energy  
are quantified by the three-point function,  
\be S_T({\bf q},t) = \frac{1}{N} 
\langle \delta f({\bf q},t) \delta e(-{\bf q},t=0)  
\rangle. 
\label{sT} 
\ee 
In Eqs.~(\ref{s4}, \ref{sT}), $\delta f({\bf q},t)$ and $\delta e({\bf q},t)$ 
denote the Fourier transforms with respect to 
${\bf x}$ of $\delta f({\bf x},t)$ and $\delta e({\bf x},t)$,  
respectively. 
We will show data for  
fixed $|{\bf k}|$, as for the dynamic susceptibilities above.  
In our numerical simulations we have also performed  
a circular averaging over wavevectors of fixed moduli $|{\bf q}|$,  
although the relative orientations of ${\bf q}$ and ${\bf k}$ 
plays a role~\cite{doliwa,Szamel-orientation}.  

It should be remarked that the spatial correlations
quantified through Eqs.~(\ref{s4}, \ref{sT}) can be measured 
in any statistical ensemble, because they are local
quantities not sensitive to far away boundary conditions. 
Therefore, their $q \to 0$ limits is related to the 
dynamic susceptibilities measured in the ensemble
where all conserved quantities fluctuate. 
%in particular density. For instance, $S_4(q\to0,t) = 
%\chi_4^{NPT}$. 
 
\begin{figure} 
\psfig{file=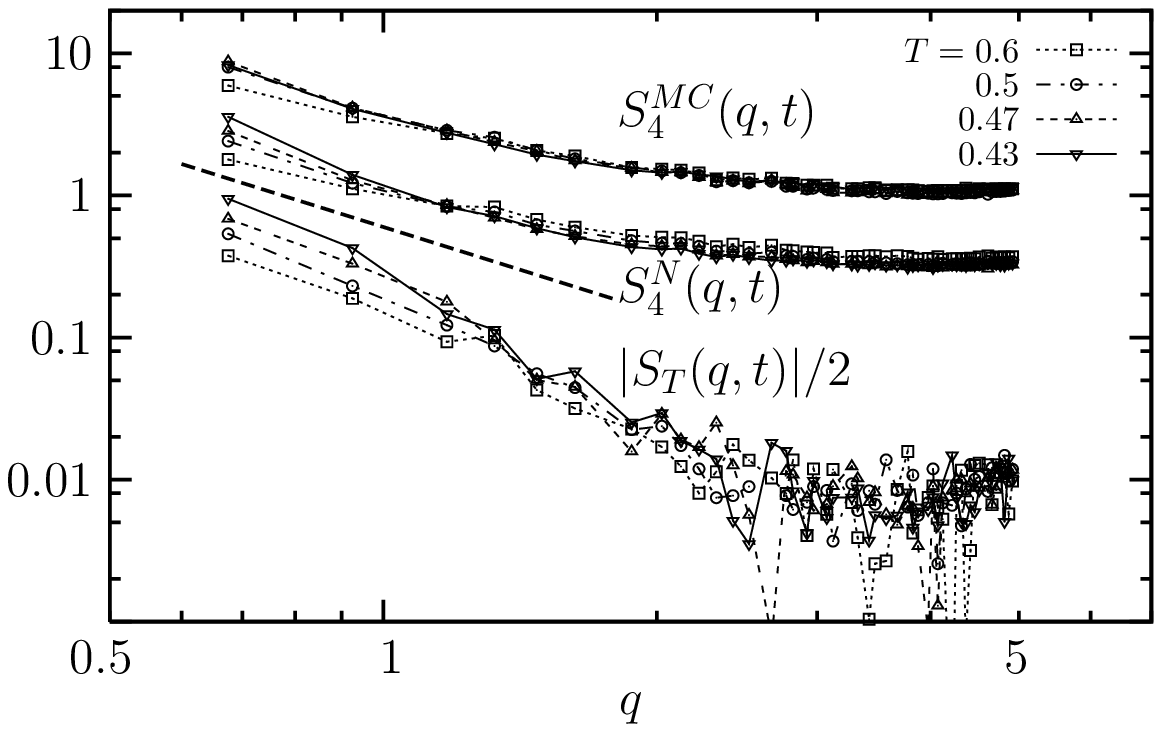,width=8.5cm} 
\psfig{file=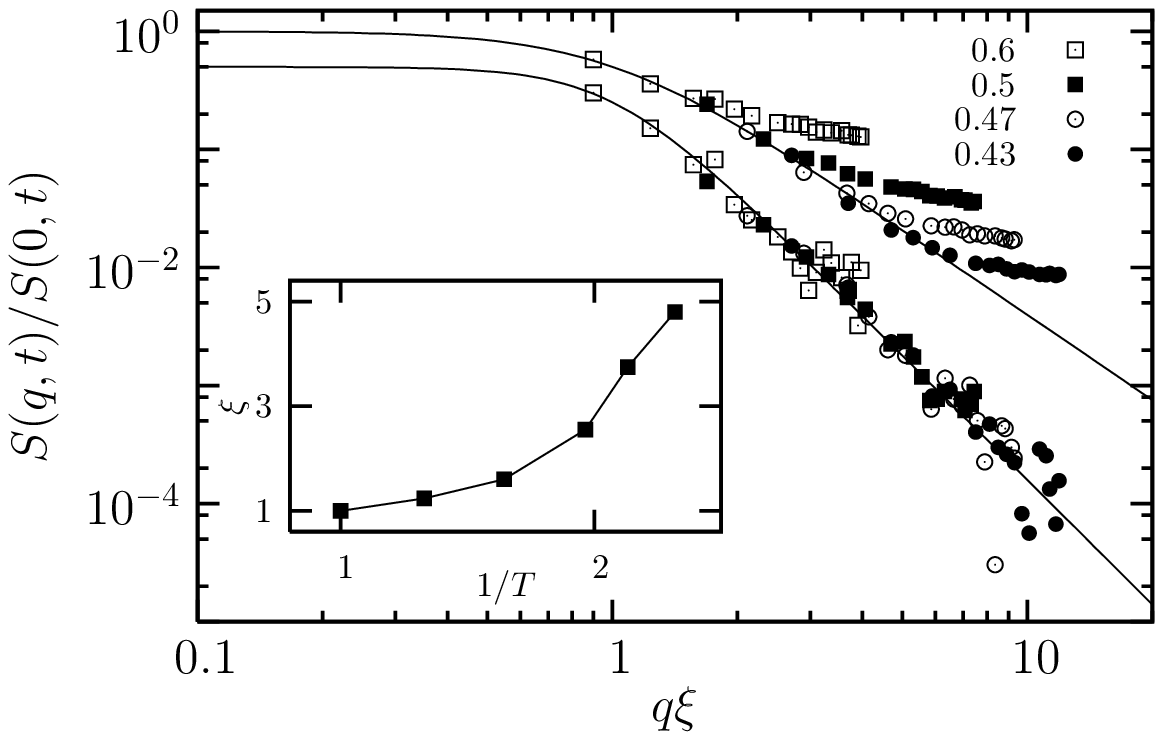,width=8.5cm} 
\caption{\label{figs}  
Top: Four-point 
dynamic structure factors from Eqs.~(\ref{s4})  
for Monte Carlo (MC) and Newtonian (N) dynamics,  
and three-point structure factor,  Eq.~(\ref{sT}), for  
Newtonian dynamics. For comparison, we show  
the power law $1/q^2$ as a dashed line. Note that $S_T$ is a negative 
quantity, so we present its absolute value. 
$S_T$ and $S_4^{MC}$ have been vertically shifted for graphical 
convenience. 
Bottom: Rescaled dynamic structure factor  
for Newtonian dynamics using Eq.~(\ref{rescales}) 
with $\alpha=2.4$ for $S_4$ and $\alpha=3.5$ for $S_T$. The same 
dynamic lengthscale $\xi_4 = \xi_T \equiv \xi$ is used in both  
cases, and the temperature evolution of $\xi$ is shown in the inset. 
} 
\end{figure} 
   
We present our numerical results for the temperature  
dependence of four-point and three-point  
structure factors in Fig.~\ref{figs}. Similar four-point  
dynamic structure factor have been discussed  
before~\cite{sharon,yamamoto,glotzer,lacevic,gc,steve,berthier,TWBBB}.  
They  
present at low $q$ a peak whose height increases while the peak position  
shifts to lower $q$ when $T$ decreases.  
This peak is unrelated to static density  
fluctuations which are small and featureless in this  
regime~\cite{KA}. 
This growing peak is direct evidence of a growing  
dynamic lengthscale, $\xi_4(T)$, associated to dynamic heterogeneity 
as temperature is decreased.  
The dynamic lengthscale $\xi_4$ should then be  
extracted from these data by fitting the  
$q$-dependence of $S_4(q,t)$ to a specific form. An  
Ornstein-Zernike form has often been used~\cite{lacevic,berthier},  
and we have presented  
its $1/q^2$ large $q$ behavior in Fig.~\ref{figs}.  
Since our primary aim is to measure dynamic susceptibilities  
on a wide range of temperatures, we have used a relatively small  
number of particles, $N=1000$. At density $\rho_0 = 1.2$, the largest  
distance we can access in spatial correlators is $L/2 \approx 5$, 
which makes an absolute determination of  
$\xi_4$ somewhat ambiguous.  
Similarly the range of wavevectors shown in Fig.~\ref{figs} is too small 
to assign a precise value even to the exponent characterizing  
the large $q$ behavior of $S_4(q,t) \sim 1/q^{\alpha}$. Our data is compatible 
with 
a value $\alpha \approx 2.4$. To extract  
$\xi_4$ we therefore fix $\alpha=2.4$  
and determine $\xi_4$ by assuming the following scaling 
behavior~\cite{TWBBB},  
\be 
S_4(q,t) = \frac{S_4(q=0,t)}{1+( q \xi_4)^\alpha},
\label{rescales} 
\ee 
using $S_4(0,t)$ and $\xi_4$ as free parameters.
The results of such an analysis  
are shown in the bottom panel of Fig.~\ref{figs}. 
This procedure leads to values for $\xi_4$  
which are in good agreement with previous determinations  
using different procedures~\cite{berthier}. In particular we find that 
a power law relationship $\xi_4 \sim \tau_\alpha^{1/z}$ with $z \approx 4.5$ 
describes our data well, as reported in Ref.~\cite{steve}  
for this system. 
 
Since dynamic structure factors probe local spatial correlations 
they do not depend on the statistical ensemble chosen 
for their calculation, at least in the thermodynamic limit. 
However, as predicted in Sec.~\ref{sectiongiulioft}, dynamic 
correlations are expected to retain a dependence on the microscopic dynamics 
of the particles, our prediction being that correlations 
should be stronger for 
Newtonian dynamics than for stochastic dynamics.  
This prediction is directly confirmed in Fig.~\ref{figs} where we  
show $S_4^{MC}(q,t)$ obtained from our Monte Carlo simulations. Clearly the  
temperature evolution of $S_4^{MC}$ is slower than that of  
$S_4^{ND}$, in agreement with the slower temperature evolution  
of $\chi_4^{MC}$ already observed in Fig.~\ref{figLJ4}. 
 
An important new result contained in Fig.~\ref{figs} is the presence
and development of a similar low-$q$ peak in the three-point structure
factor $S_T(q,t)$.  Note that, as for $\chi_T(t)$, we find that
$S_T(q,t)$ is a negative quantity. This means that a local positive
fluctuation of the energy is correlated to a local negative
fluctuation of the two-time dynamics, i.e. to a locally faster than
average dynamics. Therefore the (negative) peak in $S_T(q,t)$ is a
direct microscopic demonstration that dynamic heterogeneity is
strongly correlated to the fluctuations of at least one local
structural quantity, namely the energy~\cite{science}.  When
temperature decreases, the height of the peak in $|S_T(q,t)|$
increases and it shrinks towards lower $q$.  This is again the sign of
the presence of a second growing dynamic lengthscale, $\xi_T$, which
reflects the extent of the spatial correlations between energy and
dynamical fluctuations. Again an absolute determination of $\xi_T$ is
very hard due to system size limitations. Since we expect $\xi_T$ and
$\xi_4$ to carry equivalent physical content, we have checked that our
data are compatible with both lengthscales being equal.  In
Fig.~\ref{figs}, we rescale the three-point dynamic structure factor
using Eq.~(\ref{rescales}) with $\alpha=3.5$, and constraining $\xi_T =
\xi_4$. The scaling is of similar quality, see the bottom panel in
Fig.~\ref{figs}.  Clearly, the nontrivial $q$-dependence of $S_T(q,t)$
with a scaling collapse of reasonable quality and a length scale
consistent with that extracted from $S_4(q,t)$ is a strong indicator
that the integrated susceptibility $\chi_{T}$ grows as a result of a
unique growing length scale characteristic of dynamic heterogeneity.
A further numerical confirmation of the fact that 
the growth of the susceptibilities 
$\chi_4(t)$ and $\chi_T(t)$ cannot be attributed to an increase in the 
strength of the correlations rather than their range 
stems from the direct measurement of 
$g_4(r,t)$ and $g_T(r,t)$, the Fourier transforms of $S_4(q,t)$ and 
$S_T(q,t)$. The large distance decay of both functions can be well fitted, 
within the statistical noise, by an exponential form~\cite{doliwa} 
with a growing dynamic lengthscale but a temperature independent strength. 
Attributing all of the temperature dependence of the susceptibilities to a 
growing amplitude leads to poor fits of the spatial correlators.
This indicates that a scenario whereby the
growth of $\chi_{T}$ can be ascribed to the growth of a prefactor
with no growing length scale characteristic of dynamic heterogeneity
can be ruled out, at least for the LJ case.  Collectively, these
findings indicate that the bound for $\chi_{4}$, as first discussed
in~\cite{science}, correctly estimates a correlation volume
associated with dynamic heterogeneity.

We have carried out a similar analysis for the BKS model of 
silica~\cite{ludounpub}. Here, the analysis is
far more difficult for several reasons. First, the system is harder to
simulate than the LJ system due to the long-ranged nature of the
interactions.  Second, strong features associated with static
structure make a resolution of the low-$q$ behavior in $S_T(q,t)$
somewhat more challenging in this system.  Lastly, the overall scale
of dynamical fluctuations at the lowest temperatures studied are much
smaller than in the LJ system (see Fig.~\ref{figLJ3}).
Regardless, we do find results consistent
with a scaling scenario for $S_T(q,t)$, and, as we will see in the
following paper, the growth of $\chi_{T}$ tracks that of
$\chi_{4}^{NVE}$.  These facts give support to the notion that the
scenario for the BKS model of silica is the same as for the
LJ system although the direct supporting evidence for this is, at
this stage, not quite as strong.
 
The local correlation between energy fluctuations  
dynamic heterogeneity is broadly consistent with  
several theoretical predictions; see  the companion paper \cite{II}
for further discussion. As  
mentioned in Sec.~\ref{sectiongiulioft} 
the equality between $\xi_T$ and $\xi_4$ is a natural  
prediction, in particular close to the MCT transition. 
This is also very natural from the point  
of view of kinetically constrained models~\cite{reviewkcm}. Spin  
facilitated models, in particular,   
{\it postulate} such a correlation through the concept of 
dynamic facilitation: mobile 
sites carry positive energy fluctuations and through  
activated diffusion trigger the relaxation of  
neighboring sites~\cite{arrow}. 
In this picture a localized energy fluctuation affects the dynamics 
of a large nearby region so that there is not a 
one-to-one correspondence between slow and low-energy sites. There is  
therefore no contradiction between our results  
and the lack  of correlation between ``dynamic propensity'' 
and local potential 
energy recently reported  
in Ref.~\cite{harro}. They qualitatively agree, however, with 
recent numerical results obtained for water where a correlation
between ``dynamic'' and ``energetic'' propensities is reported~\cite{poole}.
A recent work \cite{FMV} has also suggested a relation between energy fluctuations
and finite size effects leading to a growing length at low temperature.  
 
\subsection{Summary}  
 
In this section we have discussed in detail the results  
of molecular dynamics simulations of a strong and a fragile 
glass-forming liquid. 
Our main contribution is the simultaneous  
measurement of spontaneous and induced dynamic fluctuations, and the  
quantitative confirmation in two realistic liquids  
of the central claim announced in Ref.~\cite{science}:
it is possible to obtain a quantitative estimate of the  
amplitude of dynamic fluctuations in supercooled liquids 
through the measurement of the quantity $T \chi_T/\sqrt{c_V}$, which
(once squared) gives the major contribution to $\chi_4^{NVT}$, and hence to $\chi_4^{NPT}$,
in the  low temperature regime, and is proportional to $\chi_4^{NVE}$: 
see Fig.~\ref{figLJ3}.

We have directly measured in the Lennard-Jones system  
three- and four-point dynamic structure factors  
that display slightly different wavevector dependences but  
lead nevertheless to consistent quantitative estimates of a  
dynamic correlation lengthscale, compatible with that obtained from 
$\chi_4$ and $\chi_T$. This last result is very important since 
this is direct confirmation 
that an experimental estimate of a dynamic lengthscale, 
as performed in Ref.~\cite{science}, is meaningful. 
Finally, we have found that, as predicted theoretically, 
global four-point dynamic correlations corresponding to spontaneous 
fluctuations of 
two-time  
correlators are strongly dependent on the 
microscopic dynamics, 
at variance with usual two-point correlations.
 
\section{Perspectives and conclusions} 
 
\label{conclusion} 

We conclude this rather long article, to be followed by a companion
paper~\cite{II}, with brief comments only. Four-point correlators were originally introduced to define the lengthscale of 
dynamical heterogeneities in glass formers. Our results, in that respect, are double-sided. 
We showed that global four-point functions, corresponding to the fluctuations of intensive dynamical correlators, 
not only depend on the  statistical ensemble but also on the choice of dynamics. 
The dependence on the statistical ensemble is useful to obtain lower bounds for experimentally relevant situations. 
However, on a more general ground, these dependences unveil 
that four-point correlators are more complicated than what was originally thought 
and their quantitative interpretation is somewhat flummoxed. Instead, we found that 
dynamical response functions, proxied by the temperature or density derivatives of 
two-time correlators provide a more clear and direct probe of genuine collective
dynamical effects.\\
We have given strong theoretical and numerical evidence for the most important claim made in this 
paper: the existence of unique dynamical lengthscale $\xi$ governing the growth of all the relevant dynamical susceptibilities,
independently of dynamics (and of course ensemble!). This result can be proved within the MCT of glasses, as we elaborate further 
in the companion paper~\cite{II}, but is expected more generally as soon as $\xi$ becomes somewhat large compared to 
the inter-atomic spacing. Our numerical results show that this is true both in the fragile LJ system and in the strong BKS
system: all dynamical susceptibilities ($\chi_T$, $\chi_4^{NVE}$) behave similarly, at least in the weakly supercooled region 
accessible to numerical simulations.

One rather striking result of our analysis is 
that even Arrhenius dynamics in Newtonian systems must involve some amount of dynamical correlations. This is confirmed by 
our numerical simulations on the BKS system, but the result 
holds more generally. Even a dilute assembly 
of Arrheniusly relaxing entities, e.g. 
two-level systems, should develop non-trivial 
dynamical correlations at sufficiently low temperatures, 
provided they interact with the {\it same} Newtonian thermal bath.
This is obviously the case for a strong glass-formers where 
a particle is both a relaxing entity and part of its neighbors' bath.

As for the perspective for the future, we hope that our work will 
trigger more experimental and numerical investigations of supercooled 
liquids and jamming systems, extending our results both from a quantitative and a qualitative point of view \cite{exppaperinprep}. 
In particular, the distinction between dynamical correlations (explored here) and cooperativity, if any, should be clarified.
The relation between the two notions might be very different in strong and fragile systems, and the distinction between the MCT and 
the deeply supercooled regimes might also be relevant. Is $\xi$ as defined in the present paper related to the Adam-Gibbs or 
the mosaic length scale \cite{rfot,droplet}? In this respect, the full understanding of deceivingly simple Arrhenius systems should be of great help.
      
\acknowledgments
We thank L. Cipelletti, S. Franz,
F. Ladieu, A. Lef\`evre, D. L'H{\^o}te and G. Szamel for discussions. 
We are also deeply indebted to M. Cates and G. Tarjus for insightful remarks which 
helped clarifying the manuscript.
D.R.R. and K.M. acknowledge support from the NSF (NSF CHE-0134969). 
G.B. is partially supported by EU contract HPRN-CT-2002-00307 (DYGLAGEMEM).
K.M. would like to thank J. D. Eaves for his help on 
the development of efficient numerical codes.
 
%L.B. is happy to have finished this paper
%almost 4 years after his very first $T$-derivative (16 Dec. 2002). 


\begin{thebibliography}{99} 
 
\bibitem{Donth} E. Donth, {\it The glass transition} 
(Springer, Berlin, 2001). 
 
\bibitem{DS} P.~G. Debenedetti, F.~H. Stillinger, 
Nature {\bf 410}, 259 (2001). 

\bibitem{walter} K. Binder and W. Kob,
{\it Glassy materials and disordered solids} (World Scientific, Singapore, 
2005).

\bibitem{nostructure} R. Leheny, N. Menon, S.~R. Nagel, D. L. Price, 
K. Suzuya, and P. Thiyagarajan,  J. Chem. Phys. {\bf 105}, 7783 (1996); 
A.  Tolle, H. Schober, J. Wuttke, and F. Fujara,
Phys. Rev. E, {\bf 56}, 809 (1997).
 
\bibitem{ediger} M.~D. Ediger, 
Annu. Rev. Phys. Chem. {\bf 51}, 99 (2000). 

\bibitem{sillescu} 
H. Sillescu, J. Non-Cryst. Solids {\bf 243}, 81 (1999). 
 
\bibitem{richert} 
R. Richert, J. Phys.: Condens. Matter {\bf 14}, R703 2002 . 
 
\bibitem{sharon} 
S.~C. Glotzer, J. Non-Cryst. Solids {\bf 274} 342 (2000). 

\bibitem{hans}  
H.~C. Andersen,  Proc. Natl. Acad. Sci. 
{\bf 102}, 6686 (2005). 
 
\bibitem{rfot} X.~Y. Xia and P.~G. Wolynes,  
Proc. Natl. Acad. Sci. {\bf 97}, 2990 (2000). 
 
\bibitem{Gilles} P. Viot, G. Tarjus, D. Kivelson, 
J. Chem. Phys. {\bf 112}, 10368 (2000). 
 
\bibitem{jackle} 
J. J{\"a}ckle and S. Eisinger, Z. Phys. B: Condens. Matter {\bf 84}, 115 
(1991). 
 
\bibitem{FA} G.~H. Fredrickson and H.~C. Andersen, 
Phys. Rev. Lett. {\bf 53}, 
1244 (1984); J. Chem. Phys. {\bf 83}, 958 (1984). 
 
\bibitem{harrowell}  
S. Butler and P. Harrowell, J. Chem. Phys. {\bf 95}, 4454 (1991); 
{\bf 95}, 4466 (1991); 
P. Harrowell, Phys. Rev. E {\bf 48}, 4359 (1993); 
M. Foley and P. Harrowell, J. Chem. Phys. {\bf 98}, 5069 (1993). 

\bibitem{gc} J.~P. Garrahan, D. Chandler, 
Phys. Rev. Lett. {\bf 89}, 035704 (2002). 
 
\bibitem{steve} S. Whitelam, L. Berthier, J.~P. Garrahan, 
Phys. Rev. Lett. {\bf 92}, 185705 (2004); Phys. Rev. E {\bf 71}, 026128 (2005).

\bibitem{BB} G. Biroli and J.-P. Bouchaud, 
Europhys. Lett. {\bf 67}, 21 (2004). 
 
\bibitem{doliwa} B. Doliwa and A. Heuer, 
 Phys. Rev. E {\bf 61}, 6898 (2000). 
 
\bibitem{harrowell2} 
M.~M. Hurley and P. Harrowell, Phys. Rev. E {\bf 52}, 1694 (1995); 
D.~N. Perera and P. Harrowell, J. Chem. Phys. {\bf 111}, 5441 (1999). 
 
\bibitem{yamamoto} 
R. Yamamoto and A. Onuki, Phys. Rev. Lett. {\bf 81}, 4915 (1998). 
 
\bibitem{japonais} 
Y. Hiwatari and T. Muranaka, J. Non-Cryst. Solids {\bf 235-237}, 
19 (1998). 
 
\bibitem{exp0}                           
U. Tracht, M. Wilhelm, A. Heuer, H. Feng, K. Schmidt-Rohr, and H.~W. Spiess, 
Phys. Rev. Lett. {\bf 81}, 2727 (1998). 
 
\bibitem{exp1} E. Weeks, J.C. Crocker, A.C. Levitt, 
A. Schofield, and D.A. Weitz,  
Science {\bf 287}, 627 (2000). 
 
\bibitem{exp2} 
E. Vidal-Russell,  N.~E. Israeloff, 
Nature {\bf  408}, 695 (2000). 
 
\bibitem{mark} S.~A. Reinsberg, X.~H. Qiu, M. Wilhelm, H.~W. Spiess,  
and M.~D. Ediger, J. Chem. Phys. {\bf 114}, 7299 (2001). 
 
\bibitem{encoremark} X.~H. Qiu and M.~D. Ediger, 
J. Phys. Chem. B {\bf 107}, 459 (2003). 
 
\bibitem{encoredonth} 
E. Hempel, G. Hempel, A. Hensel, C. Schick, and E. Donth, 
J. Phys. Chem. B {\bf 104}, 2460 (2000). 
 
\bibitem{FP} S. Franz, G. Parisi, 
J. Phys.: Condens. Matter {\bf 12}, 6335 (2000). 
 
\bibitem{parisi} 
G. Parisi, J. Phys. Chem. B {\bf 103}, 4128 (1999). 
 
\bibitem{silvio2} 
S. Franz, C. Donati, G. Parisi, and S.~C. Glotzer,  
Phil. Mag. B, {\bf 79}, 1827,  
(1999); 
C. Donati, S. Franz, S.~C. Glotzer, and G. Parisi,  
J. Non-Cryst. Solids {\bf 307}, 215 (2002).  
 
\bibitem{glotzer} 
C. Bennemann, C. Donati, J. Baschnagel, S.~C. Glotzer, 
Nature {\bf 399}, 246 (1999). 
 
\bibitem{lacevic} 
N. La\v{c}evi\'{c}, F.~W. Starr, T.~B. Schroder, and S.~C. Glotzer,  
J.Chem.Phys. {\bf 119}, 7372 (2003). 

\bibitem{berthier} L. Berthier, 
Phys. Rev. E {\bf  69}, 020201 (2004). 
 
\bibitem{TWBBB} C. Toninelli, M. Wyart, G. Biroli, L. Berthier, 
J.-P. Bouchaud, Phys. Rev. E {\bf 71}, 041505 (2005). 
 
\bibitem{mayer}  
P. Mayer, H. Bissig, L. Berthier, L. Cipelletti, J.~P. Garrahan, P. 
Sollich, and V. Trappe, 
Phys. Rev. Lett. {\bf 93}, 115701 (2004). 
 
\bibitem{dauchot} O. Dauchot, G. Marty and G. Biroli,
Phys. Rev. Lett. {\bf 95}, 265701, (2005). 
 
\bibitem{science} L. Berthier, G. Biroli, J.-P. Bouchaud, L. Cipelletti,  
D. El Masri, D. L'H{\^o}te, F. Ladieu, and M. Pierno, 
Science {\bf 310},  
1797 (2005). 
 
\bibitem{II} 
L. Berthier, G. Biroli, J.-P. Bouchaud,  
W. Kob, K. Miyazaki, and D.~R. Reichman, companion paper,
cond-mat/0609658. 

 \bibitem{BBMR} G. Biroli, J.-P. Bouchaud, K. Miyazaki, 
and D.~R. Reichman, Phys. Rev. Lett. {\bf 97}, 195701 (2006).
 
\bibitem{gleim} T. Gleim, W. Kob, and K. Binder, 
 Phys. Rev. Lett. {\bf 81}, 004404 (1998).  
 
\bibitem{tobepisa} L. Berthier and W. Kob, 
to be published in J. Phys. Condens. Matter, condmat/0610253.

\bibitem{leticia} 
J.-P. Bouchaud, L.~F. Cugliandolo, J. Kurchan and M. M{\'e}zard, 
Physica 
A {\bf 226}, 243 (1996). 

\bibitem{Gotze} W. G{\"o}tze, J. Phys. Cond. Matt. {\bf 11}, A1 (1999). 
W. G{\"o}tze and L. Sj{\"o}gren, Rep. Prog. Phys. {\bf 55}, 241 (1992). 
 
\bibitem{reviewkcm} 
F. Ritort and P. Sollich, Adv. Phys. {\bf 52}, 219 (2003). 

\bibitem{exppaperinprep} C. Alba-Simioneso, 
L. Berthier, G. Biroli, J.-P. Bouchaud, C. Dalle-Ferrier, D. L'H{\^o}te, 
F. Ladieu, G. Tarjus, C. Thibierge, in preparation.
 
\bibitem{Ernst1991} 
R.~M. Ernst, S.~R. Nagel, and G.~S. Grest,  
Phys. Rev. B {\bf 43}, 8070, (1991).  
 
\bibitem{young} 
K. Binder and A.~P. Young, Rev. Mod. Phys. {\bf 58}, 801 (1986). 
 
\bibitem{ea}  
S.~F. Edwards and P.~W. Anderson, 
J. Phys. F : Metal Phys. {\bf 5} 965 (1975). 

\bibitem{nonlin} J.-P. Bouchaud, G. Biroli, 
Phys. Rev. B {\bf 72} 064204 (2005).
 
\bibitem{MS0} This was recently shown rigourously for a 
large class of glassy dynamics in 
A. Montanari, G. Semerjian, J. Stat. Phys. {\bf 125}, 22 (2006),
and suggested in full
generality by the bound derived in \cite{science}.

\bibitem{hansen} J.~P. Hansen, I.~R. Mc Donald,  
{\it Theory of simple liquids} (Elsevier, Amsterdam, 1986). 

\bibitem{KT}  T.~R. Kirkpatrick and D. Thirumalai, 
Phys. Rev. A {\bf 37}, 4439 (1988). 

\bibitem{Ck} 
D. Chandler, J.~P. Garrahan, R.~L. Jack, L. Maibaum and A.~C. Pan, 
Phys. Rev. E {\bf 74}, 051501 (2006).
 
\bibitem{luca} 
L. Cipelletti and L. Ramos, J. Phys. : Condens. Matter {\bf 17}, R253 (2005). 
 
\bibitem{at} M. Allen and D. Tildesley,  
{\it Computer Simulation of Liquids} (Oxford University Press, Oxford, 1987).  
 
\bibitem{dh} B. Doliwa and A. Heuer,  
Phys. Rev. E {\bf 61}, 6898 (2000). 
 
\bibitem{pusey} P.~N. Pusey, W. van Megen,  
Nature {\bf 320}, 340 (1986).  
 
\bibitem{lebo} J.~L. Lebowitz, J.~K. Percus, L. Verlet,  
Phys. Rev. {\bf 153}, 250 (1967). 
 
\bibitem{edigertheo} M.~D. Ediger, 
J. Non-Cryst. Solids {\bf 235-237}, 10 (1998).

\bibitem{landau} 
L.~D. Landau and E.~M. Lifshitz, {\it Statistical Physics,  
Course of Theoretical Physics, Vol. 5, Pt. 1} (Pergamon, New York, 1980). 
 
\bibitem{alba} G. Tarjus, D. Kivelson, S. Mossa, C. Alba-Simionesco,  
J. Chem. Phys. {\bf 120}, 6135 (2004). 

\bibitem{Zinn-Justin} J. Zinn Justin, {\it Quantum field  
theory and critical phenomena} (Oxford University Press, Oxford, 2002). 

\bibitem{Dean}  
D.~S. Dean, J. Phys. A {\bf 29}, L613 (1996). 
 
\bibitem{Kawa}  
K. Kawasaki, Phyisica A {\bf 208}, 35 (1994). 
 
\bibitem{ABL} A. Andreanov, G. Biroli, and A. Lef{\`e}vre, 
J. Stat. Mech. P07008 (2006). 
 
\bibitem{das} S.~P. Das, Rev. Mod. Phys. {\bf 76}, 785 (2004). 
 
\bibitem{Cardy} J. Cardy, {\it Scaling and Renormalization in Statistical  
Physics} (Cambridge University Press, Cambridge, 1996). 
 
\bibitem{DeDominicisMartin} C. De Dominicis and P. Martin, 
J. Math. Phys. {\bf 5}, 14 and 31 (1964). 
 
\bibitem{BlaizotRipka}
J.-P. Blaizot, G. Ripka,
{\it Quantum Theory of Finite Systems} (Editions Phenix, Kiev, 1998).

\bibitem{kunidave} K. Miyazaki and D. R. Reichman,
J. Phys. A: Math. Gen. {\bf 38}, L343 (2005).

\bibitem{szamel2} 
G. Szamel, and E. Flenner,
Europhys. Lett. {\bf 67}, 779 (2004). 
 
\bibitem{KA} W. Kob and H.~C. Andersen, 
Phys. Rev. Lett. {\bf 73}, 1376 (1994); 
Phys. Rev. E {\bf 53}, 4134 (1995); Phys. Rev. E {\bf 51}, 4626 (1995). 
 

\bibitem{beest90} 
B.~W.~H. van Beest, G.~J. Kramer, and R.~A. van Santen, 
Phys. Rev. Lett. {\bf 64}, 1955 (1990). 


\bibitem{bks_sim} 
S.~N. Taraskin and S.~R. Elliott, 
Europhys. Lett. {\bf 39}, 37 (1997); 
M. Benoit, S. Ispas, P. Jund, and R. Jullien, 
Eur. Phys. J. B {\bf 13}, 631 (2000); 
J. Horbach and W. kob, Phys. Rev. B {\bf 60}, 3169 (1999); 
Phys. Rev. E {\bf 64}, 041503 (2001).  

\bibitem{unpub} J. Horbach and W. Kob, unpublished.

\bibitem{ludounpub} L. Berthier, unpublished.
 
\bibitem{vogel} 
M. Vogel and S.~C. Glotzer, 
Phys. Rev. Lett. {\bf 92}, 255901 (2004);  
Phys. Rev. E {\bf 70}, 061504 (2004). 

\bibitem{fss} L. Berthier, 
Phys. Rev. Lett. {\bf 91}, 055701 (2003).  
 
\bibitem{szamel} G. Szamel and E. Flenner,
Phys. Rev. E {\bf 74}, 021507 (2006). 

\bibitem{Szamel-orientation}
E. Flenner, G. Szamel, cond-mat/0608398.

\bibitem{arrow} 
J.~P. Garrahan and D. Chandler, 
Proc. Natl. Acad. Sci. {\bf 100}, 9710 (2003);
L. Berthier and J.~P.~Garrahan, Phys. Rev. E {\bf 68}, 041201 (2003). 
 
\bibitem{harro}  
A. Widmer-Cooper, P. Harrowell, and H. Fynewever, 
Phys. Rev. Lett. {\bf 93}, 135701 (2004);  
A. Widmer-Cooper and P. Harrowell, 
Phys. Rev. Lett. {\bf 96}, 185701 (2006).  

\bibitem{sri}
S. Sastry, Phys. Rev. Lett. {\bf 85}, 590 (2000).

\bibitem{poole}
G.~S. Matharoo, M.~S.~G. Razul, and P.~H. Poole, 
Phys. Rev. E {\bf 74}, 050502 (2006).

\bibitem{FMV} L.A. Fernandez, V. Martin-Mayor and P. Verrocchio
Phys. Rev. E {\bf 73}, 020501 (2006).

\bibitem{droplet}
J.-P. Bouchaud and G. Biroli, J. Chem. Phys. {\bf 121}, 7347 (2004). 
 
\end{thebibliography}
\end{document}